\documentclass[aps,prb,twocolumn,showpacs,amsmath,amssymb]{revtex4}
\usepackage{dcolumn}
\usepackage{bm}
\usepackage{graphicx}
\usepackage{subfigure}
\usepackage{epstopdf}
\usepackage[makeroom]{cancel}

\begin{document}
\title{Spin-polarized edge currents and Majorana fermions in one- and two-dimensional topological superconductors}
\author{Kristofer Bj\"{o}rnson$^1$, Sergey. S. Pershoguba$^2$, Alexander. V. Balatsky$^{2,3}$, and Annica M. Black-Schaffer$^1$}
\affiliation{$^1$Department of Physics and Astronomy, Uppsala University, Box 516, S-751 20 Uppsala, Sweden}
\affiliation{$^2$Nordita, Center for Quantum Materials, KTH Royal Institute of Technology, and Stockholm University, Roslagstullsbacken 23, S-106 91 Stockholm, Sweden}
\affiliation{$^3$Institute for Materials Science, Los Alamos National Laboratory, Los Alamos, NM 87545, USA}

\begin{abstract}
We investigate the persistent currents, spin-polarized local density of states, and spectral functions of topological superconductors constructed by placing ferromagnetic impurities on top of an $s$-wave superconductor with Rashba spin-orbit interaction.
We solve self-consistently for the superconducting order parameter and investigate both two-dimensional blocks and one-dimensional wires of ferromagnetic impurities, with the magnetic moments pointing both perpendicular and parallel to the surface.
We find that the topologically protected edge states of ferromagnetic blocks give rise to spin-polarized edge currents, but that the total persistent current flows in opposite direction to what is expected from the dispersion relation of the edge states. 
We also show that the Majorana fermions at the end points of one-dimensional wires are spin-polarized, which can be directly related to the spin-polarization of the edge currents of two-dimensional blocks. Connections are also made to the physics of the Yu-Shiba-Rusinov states for zero-dimensional impurities.
\end{abstract}

\maketitle

\section{Introduction}
During the past one and a half decade, topology has come to play an important role in several different but related aspects of condensed matter physics.
In 2001, Kitaev proposed that a one-dimensional (1D) spinless $p$-wave superconductor can provide a route towards topological quantum computation, since it can host topologically protected Majorana fermions.\cite{PhysUsp.44.131}
A few years later the subject of topological insulators was born, with the essential feature being an insulating bulk and topologically protected metallic interface states inside the bulk gap.\cite{PhysRevLett.95.226801, PhysRevLett.96.106802, Science.314.1757, Science.318.766, Nature.452.970, NatPhys.5.398, NatPhys.5.438, Science.325.178}
The close connection between Majorana fermions and the edge states in topological insulators was explicitly made clear through the application of topological band theory to superconductors.\cite{RevModPhys.82.3045, RevModPhys.83.1057}
An early proposal of such a topological superconductor involved interfacing a topological insulator with a conventional $s$-wave superconductor.\cite{PhysRevLett.100.096407}
This was followed by proposals to construct a related setup using more conventional building blocks, such as $s$-wave superconductivity, magnetism, and Rashba spin-orbit interaction.\cite{PhysRevLett.103.020401, PhysRevB.82.134521, PhysRevLett.104.040502, PhysRevB.81.125318, PhysRevLett.105.077001, PhysRevLett.105.177002}
This latter approach is particularly interesting because it can physically be realized in many different ways and therefore leaves plenty of room for experimental implementation.

One of the most recent experimental investigations of Majorana fermions involves putting ferromagnetic impurities on top of a superconductor to create a 1D topological superconductor wire, where surface effects give rise to the needed Rashba spin-orbit interaction.\cite{Science.346.602, arXiv.1505.06078, arXiv.1507.03104}
The same setup can also be used to assemble structures such as 2D ferromagnetic islands or blocks, which have attracted attention because of their topologically protected edge states.\cite{PhysRevLett.114.236803, arXiv.1501.00999}
With this particular approach in mind, but also knowing that the system can be implemented in a variety of ways,\cite{Science.336.1003, NatPhys.8.795, NatPhys.8.887} we here investigate both block and wire configurations of magnetic moments on top of a Rashba spin-orbit coupled superconductor. To accurately model these systems, we solve self-consistently for the superconducting order parameter.

In particular, we investigate the persistent currents, local density of states (LDOS), and the spectral functions.
We recently investigated the current patterns for a magnetic point impurity,\cite{arXiv:1505.01672} providing insight into the properties of the individual atoms in ferromagnetic domains and also connecting to the well-known physics of Yu-Shiba-Rusinov states of magnetic impurities in superconductors.
Here, we instead start from the topological band theory of a 2D ferromagnetic block and investigate the resulting edge states, which are shown to be spin-polarized.
We then study 1D wires and connect their properties to the 2D blocks, by considering them as blocks with a width of a single site.
Specifically, the end-point Majorana fermions are shown to be spin-polarized with a spin-polarization directly related to that of the edge states of the block.
Through this approach we are able to fully elucidate the close connection between the topological states in 2D and 1D, as well as the connection to the Yu-Shiba-Rusinov states in 0D. In addition, we find persistent currents around the ferromagnetic block, which surprisingly flows in opposite direction to what is expected from the dispersion relation of the topologically protected edge states.
This apparent contradiction is resolved as the gap-crossing edge states are indeed found to give rise to a current compatible with their dispersion relation, but lower energy quasiparticles are found to produce even stronger counter-propagating currents, a phenomenon known in other superconductors.\cite{PhysRevB.84.214509, PhysRevB.88.184506}
This has important experimental consequences, since physical probes such as superconducting quantum interference devices (SQUIDs), which measure the magnetic field generated from the total persistent currents, will give opposite results to probes measuring transport properties involving only states close to the Fermi level.

\section{Model}
Motivated by the presence of Rashba spin-orbit interaction, $s$-wave superconductivity, and magnetic impurities in recent experimental setups,\cite{Science.346.602, arXiv.1505.06078, arXiv.1507.03104} we consider the following Hamiltonian
\begin{align}
	\mathcal{H} &= \mathcal{H}_{kin} + \mathcal{H}_{V_z} + \mathcal{H}_{so} + \mathcal{H}_{sc},
\label{Equation:Tight_binding_Hamiltonian} \\ 
	\mathcal{H}_{kin} &= -t\sum_{\langle\mathbf{i},\mathbf{j}\rangle,\sigma}c_{\mathbf{i}\sigma}^{\dagger}c_{\mathbf{j}\sigma} - \mu\sum_{\mathbf{i},\sigma}c_{\mathbf{i}\sigma}^{\dagger}c_{\mathbf{i}\sigma}, \nonumber \\ 
	\mathcal{H}_{V_z} &= -\sum_{\mathbf{i},\sigma,\sigma'}\left(V_z(\mathbf{i})\hat{\mathbf{n}}\cdot\boldsymbol{\sigma}\right)_{\sigma\sigma'}c_{\mathbf{i}\sigma}^{\dagger}c_{\mathbf{i}\sigma'}, \nonumber \\ 
	\mathcal{H}_{so} &= \frac{\alpha}{2}\sum_{\mathbf{i}\mathbf{b}}\left(e^{i\theta_{\mathbf{b}}}c_{\mathbf{i}+\mathbf{b}\downarrow}^{\dagger}c_{\mathbf{i}\uparrow} + {\rm H.c.}\right), \nonumber \\ 
	\mathcal{H}_{sc} &= \sum_{\mathbf{i}}\left(\Delta_{\mathbf{i}}c_{\mathbf{i}\uparrow}^{\dagger}c_{\mathbf{i}\downarrow}^{\dagger} + {\rm H.c.}\right) \nonumber.
\end{align}
Here $\mathbf{i}$ and $\mathbf{j}$ are site indices on a square lattice, $\mathbf{b}$ is a vector pointing along the four types of nearest neighbor bonds, $\theta_{\mathbf{b}}$ is the angular coordinate of $\mathbf{b}$, and $\boldsymbol{\sigma}$ is the vector of Pauli matrices.
The parameters of the model are the strength of the hopping parameter $t$, chemical potential $\mu$, Rashba spin-orbit interaction $\alpha$, and superconducting pair potential $V_{sc}$ of the superconducting substrate, as well as the effective Zeeman splitting coming from the ferromagnetic impurities $V_z(\mathbf{i})$.
The superconducting pair potential is used to self-consistently determine the order parameter $\Delta_{\mathbf{i}} = -V_{sc}\langle c_{\mathbf{i}\downarrow}c_{\mathbf{i}\uparrow}\rangle = -V_{sc}\sum_{E_{\nu}<0}v_{\nu\mathbf{i}\downarrow}^{*}u_{\mathbf{i}\uparrow}$,
where $u_{\nu\mathbf{i}\sigma}$ and $v_{\nu\mathbf{i}\sigma}$ are the $\sigma$-spin electron and hole components, respectively, of the $\nu$th eigenstate.
We here set the hopping parameter $t = 1$, Rashba spin-orbit interaction $\alpha = 0.56$, chemical potential $\mu = -4$, and superconducting pair potential $V_{sc} = 5.36$. The specific values are not important, but our results are generally valid in a wide range of parameter space.\cite{PhysRevB.88.024501}
The Zeeman term is set to zero everywhere, except for a block- or line-shaped region, where it is set to a finite $V_z$, with the direction determined by the unit vector $\hat{\mathbf{n}}$.
This allows us to study both 2D blocks and 1D wires of ferromagnetic impurities embedded within a conventional superconductor.

\subsection{Currents}
The most essential property of topologically nontrivial systems is the appearance of gapless edge states.
In the case of quantum Hall systems, these give rise to persistent currents along the edges, while for topological insulators they instead result in persistent spin-currents.
Likewise, topological superconductors have gapless edge states, and it is therefore of interest to investigate the persistent currents in these systems.
In fact, we find that a study of these currents help us understand also the spin-polarization of Majorana fermions at the end points of 1D wires.
We thus need not only expressions for calculating currents, but the spin-polarized currents for any given spin-polarization axis.

To derive expressions for the currents, we consider the time rate of change of the spin-density operator $\hat{\rho}_{\mathbf{i}\sigma} = c_{\mathbf{i}\sigma}^{\dagger}c_{\mathbf{i}\sigma}$:
\begin{align}
	\label{Equation:Spin_density_time_rate_of_change}
	\frac{d\hat{\rho}_{\mathbf{i}\sigma}}{dt} =& \frac{i}{\hbar}\left[\mathcal{H}, \hat{\rho}_{\mathbf{i}\sigma}\right].
\end{align}
Because we are interested in arbitrary spin directions, the operators have to be understood to be written in the basis of the particular spin direction of interest.
If $\sigma$ is in the direction $(\theta,\varphi)$ in spherical coordinates, the operators are related to the operators in the basis of the Hamiltonian according to
\begin{align}
	c_{\mathbf{i}\sigma}^{\dagger} = \cos\left(\frac{\theta}{2}\right)c_{\mathbf{i}\uparrow}^{\dagger} + \sin\left(\frac{\theta}{2}\right)e^{i\varphi}c_{\mathbf{i}\downarrow}^{\dagger}.
\end{align}
Now let $\bar{\sigma}$ denote the opposite spin direction to $\sigma$ and define
\begin{align}
	\label{Equation:Source_and_current_operators}
	\hat{S}_{\mathbf{i}\sigma} =& a_{\mathbf{i}\bar{\sigma}\sigma}c_{\mathbf{i}\sigma}^{\dagger} c_{\mathbf{i}\bar{\sigma}} + a_{\mathbf{i}\bar{\sigma}\sigma}^{*}c_{\mathbf{i}\bar{\sigma}}^{\dagger} c_{\mathbf{i}\sigma},\nonumber\\
	\hat{J}_{\mathbf{i}\mathbf{b}\sigma\sigma} =& -\left(a_{\mathbf{i}\mathbf{b}\sigma\sigma}c_{\mathbf{i}+\mathbf{b}\sigma}^{\dagger} c_{\mathbf{i}\sigma} + a_{\mathbf{i}\mathbf{b}\sigma\sigma}^{*}c_{\mathbf{i}\sigma}^{\dagger} c_{\mathbf{i}+\mathbf{b}\sigma}\right),\nonumber\\
	\hat{J}_{\mathbf{i}\mathbf{b}\sigma\bar{\sigma}} =& -\left(a_{\mathbf{i}\mathbf{b}\bar{\sigma}\sigma} c_{\mathbf{i}+\mathbf{b}\bar{\sigma}}^{\dagger}c_{\mathbf{i}\sigma} + a_{\mathbf{i}\mathbf{b}\bar{\sigma}\sigma}^{*}c_{\mathbf{i}\sigma}^{\dagger} c_{\mathbf{i}+\mathbf{b}\bar{\sigma}}\right),
\end{align}
where $a_{\mathbf{i}\bar{\sigma}\sigma}$, $a_{\mathbf{i}\mathbf{b}\sigma\sigma}$, and $a_{\mathbf{i}\mathbf{b}\bar{\sigma}\sigma}$ are coefficients to be determined later.
It is then possible to show, as is done in Appendix \ref{Appendix:Derivation_of_expressions_for_currents}, that Eq.~\eqref{Equation:Spin_density_time_rate_of_change} can be written as
\begin{align}
	\label{Equation:Spin_density_time_rate_of_change_source_current_operators}
	\frac{d\hat{\rho}_{\mathbf{i}\sigma}}{dt} =& \hat{S}_{\mathbf{i}\sigma} - \sum_{\mathbf{b}}\left(\hat{J}_{\mathbf{i}\mathbf{b}\sigma\sigma} + \hat{J}_{\mathbf{i}\mathbf{b}\sigma\bar{\sigma}}\right),
\end{align}
where the sum runs over all bonds $\mathbf{b}$ for which there is a term in the Hamiltonian (for Eq.~\eqref{Equation:Tight_binding_Hamiltonian} only nearest neighbor bonds).
It is clear from Eq.~\eqref{Equation:Source_and_current_operators} that $\hat{S}_{\mathbf{i}\sigma}$ is an on-site source operator which converts $\bar{\sigma}$-spins into $\sigma$-spins.
This operator is unrelated to currents and is thus of no further interest here.
Next, $\hat{J}_{\mathbf{i}\mathbf{b}\sigma\sigma}$ is a current operator for $\sigma$-spins moving away from site $\mathbf{i}$ along bond $\mathbf{b}$, and is from Eq.~\eqref{Equation:Spin_density_time_rate_of_change_source_current_operators} seen to be responsible for decreasing the number of $\sigma$-spins on site $\mathbf{i}$.
Finally, also $\hat{J}_{\mathbf{i}\mathbf{b}\sigma\bar{\sigma}}$ is responsible for transferring $\sigma$-spins away from site $\mathbf{i}$ along bond $\mathbf{b}$, but it flips the spin in the process.

While $\hat{J}_{\mathbf{i}\mathbf{b}\sigma\sigma}$ allows us to investigate spin-polarized currents, the total current is obtained by adding all contributions to the current and we are therefore also interested in the operator $\hat{J}_{\mathbf{i}\mathbf{b}} = \sum_{\sigma\sigma'}\hat{J}_{\mathbf{i}\mathbf{b}\sigma\sigma'}$.
Here the summation over $\sigma'$ ensures that we add both spin-polarized currents as well as 'spin-flipping' currents, while the summation over $\sigma$ ensures that we consider the outflow of both types of spins.
To be able to represent the current as a vector field, we also construct the on-site current operator by adding the currents along the bonds away from site $\mathbf{i}$ into the directed-average vector operator $\hat{\mathbf{J}}_{\mathbf{i}} = \frac{1}{2}\sum_{\mathbf{b}}\mathbf{b}\hat{J}_{\mathbf{i}\mathbf{b}}$,
and similarly for the spin-polarized and spin-flipping currents.

To fully construct the current operators we also need to evaluate the $a$-coefficients in Eq.~\eqref{Equation:Source_and_current_operators}.
This is complicated by the fact that we are interested in spin-polarized currents for arbitrary spin-polarization axes.
In addition, the final expressions for the currents are obtained by evaluating expectation values $\langle\hat{\mathbf{J}}_{\mathbf{i}}\rangle$ and $\langle\hat{\mathbf{J}}_{\mathbf{i}\sigma\sigma'}\rangle$, where the calculation also needs to be done in the particular basis.
A detailed derivation of the method used to calculate currents is provided in Appendix \ref{Appendix:Derivation_of_expressions_for_currents}.
In the following we present the current in two complementary ways.
First of all, $\max_{\mathbf{i}}|\langle\hat{\mathbf{J}}_{\mathbf{i}}\rangle|$ as a function of the magnetic moment allows us to identify the onset of persistent currents in the material.
However, because the maximum is taken over the whole sample, it gives no information about where these currents appear.
Therefore, for selected parameters we also show the current vector field.

\subsection{Spin-polarized LDOS}
For 1D wires the topologically protected edge states manifest themselves in the form of Majorana fermions.
For this reason we are interested in calculating the LDOS for a cut along the wire.
As we want to relate the spin-polarization of these Majorana fermions to the spin-polarization of the edge currents of 2D blocks, we are further interested in the spin-polarized LDOS for a given spin-polarization axis.
This is calculated using the expression
\begin{widetext}
\begin{align}
	\rho(\mathbf{i}, \theta, \varphi, E) = \sum_{\nu}\left[\begin{array}{cc}
			u_{\nu\mathbf{i}\uparrow}^{*} & u_{\nu\mathbf{i}\downarrow}^{*}
		\end{array}\right]\left[\begin{array}{cc}
			\cos^2(\frac{\theta}{2}) & \cos(\frac{\theta}{2})\sin(\frac{\theta}{2})e^{-i\varphi}\\
			\cos(\frac{\theta}{2})\sin(\frac{\theta}{2})e^{i\varphi} & \sin^2(\frac{\theta}{2})
		\end{array}\right]\left[\begin{array}{c}
			u_{\nu\mathbf{i}\uparrow}\\
			u_{\nu\mathbf{i}\downarrow}
		\end{array}\right]\delta(E - E_{\nu}).
\end{align}
\end{widetext}
The total LDOS is obtained by adding the spin-polarized LDOS for two opposite spin-polarizations.

\subsection{Spectral function}
The edge states of the 2D block also appear in the spectral function.
Due to the limited size of the simulated system it is important to single out the contribution to the spectral function coming from the edge states, otherwise bulk states pollute the edge spectrum.
We do this by introducing a state classification function $C(\nu)$, which is one for every state classified as an edge state, and zero otherwise.
We here define a state to be an edge state if more than $50\%$ of its probability amplitude is located inside a region, six sites wide and symmetrically positioned across the boundary. With this definition, the contribution to the spectral function coming from edge states can be calculated as
\begin{align}
	\label{Equation:Spectral_function}
	A(\mathbf{k}, E) =& \sum_{\nu,\sigma}|u_{\nu k_x\sigma}|^2\delta(E - E_{\nu})C(\nu),
\end{align}
where $u_{\nu k_x\sigma}$ is the Fourier transform of $u_{\nu x\sigma}$, and $u_{\nu x\sigma}$ is the restriction of $u_{\nu\mathbf{i}\sigma}$ to the $y$-coordinate at which the bottom of the ferromagnetic block is located.

\section{Dimensional reduction}
\label{Section:Dimensional_reduction}
For a homogeneous bulk it is possible to write the 2D model of Eq.~\eqref{Equation:Tight_binding_Hamiltonian} as\cite{PhysRevB.82.134521}
\begin{align}
	\mathcal{H}(\mathbf{k}) =& \left[\begin{array}{cc}
		H_{0}(\mathbf{k}) & \Delta(\mathbf{k})\\
		\Delta^{\dagger}(\mathbf{k}) & -H_{0}^{T}(-\mathbf{k})
	\end{array}\right],
\end{align}
where
\begin{align}
	\label{Equation:Bulk_Hamiltonian}
	H_{0}(\mathbf{k}) =& \epsilon(\mathbf{k}) - V_z\sigma_z - \mathcal{L}_{0}(\mathbf{k})\cdot\boldsymbol{\sigma},\\
	\Delta(\mathbf{k}) =& i\Delta\sigma_y,\\
	\epsilon(\mathbf{k}) =& -2t\left[\cos(k_x) + \cos(k_y)\right] - \mu,\\
	\mathcal{L}_{0}(\mathbf{k}) =& \alpha(\sin(k_y), -\sin(k_x), 0).
\end{align}
This Hamiltonian have been topologically classified and the relevant condition for being in a topologically nontrivial phase is\cite{PhysRevB.82.134521, PhysRevB.88.024501}
\begin{align}
	\label{Equation:Topological_phase_transition_condition}
	0 < |\Delta|^2 < V_z^2 - (4t+\mu)^2 = V_z^2,
\end{align}
where the last equality follows because $\mu=-4t$.
Moreover, making the restriction $k_y = 0$ in Eq.~\eqref{Equation:Bulk_Hamiltonian} and defining $\tilde{\mu} = \mu + 2t$, we obtain
\begin{align}
	\label{Equation:Bulk_Hamiltonian_1D}
	H_{0}^{1D}(\mathbf{k}) =& -2t\cos(k_x) - \tilde{\mu} - V_z\sigma_z + \alpha\sin(k_x)\sigma_y,
\end{align}
which describes a 1D superconducting wire oriented along the $x$-direction, recently intensively studied in e.g.~Refs.~[\onlinecite{PhysRevLett.105.077001, PhysRevLett.105.177002, PhysRevB.89.134518}].
It can easily be confirmed that this model, just like the 2D case, goes through a topological phase transition at $|\Delta|^2 = V_z^2 - (2t+\tilde{\mu})^2$.
This is one example of a dimensional reduction from two to one dimension, and similar relations play an important role in this work.

Note, however, that if the 1D wire is embedded in a 2D surface, this type of dimensional reduction makes little sense physically.
First of all, setting $k_y = 0$ corresponds to considering wave-functions that are completely delocalized in the $y$-direction.
Second, a 1D infinite wire is produced by setting $V_z(\mathbf{i}) = V_z\delta_{y}$, not by ignoring the $y$-direction.
A proper dimensional reduction would rather involve integrating out the $y$-dependence to arrive at an equation of similar form as Eq.~\eqref{Equation:Bulk_Hamiltonian_1D}. 
This would potentially renormalize the parameters, and the condition for being in the nontrivial phase should then be understood in terms of these renormalized parameters.
We explicitly point this out here, because, while we have the strict condition in Eq.~\eqref{Equation:Topological_phase_transition_condition} to rely on for identifying the topological phase transition of a 2D geometry, we have no such condition for a 1D wire.
Instead, we have to rely on a range of consequences, such as the appearance of Majorana fermions and other indicators related to the current, in order to identify the nontrivial phase of the wire.

\section{Results}
\subsection{Block with magnetic moments perpendicular to surface}
\begin{figure}
\includegraphics[width=245pt]{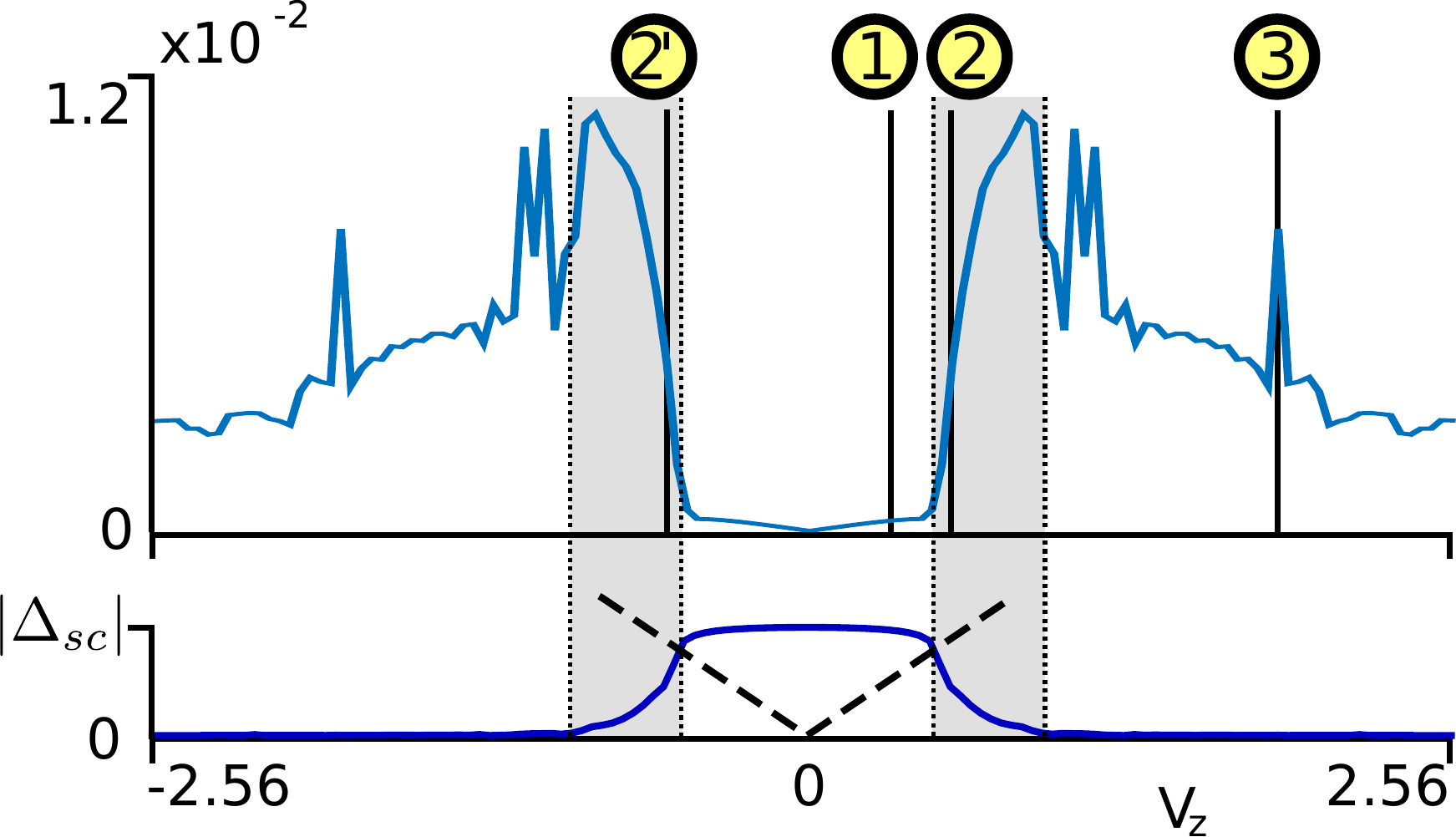}
\caption{(Color online.) Bottom: Order parameter in the center of the ferromagnetic block (blue line) as a function of the Zeeman term $V_z$ in the direction $\hat{\mathbf{n}} = \hat{\mathbf{z}}$.
Absolute value of the Zeeman term is indicated with dashed lines.
Shaded regions indicate the topologically nontrivial phase where $|V_z| > |\Delta| > 0$.
Top: Maximum current density as a function of the Zeeman term.}
\label{Figure:D_Jmax_block_perp}
\end{figure}

\begin{figure}
\includegraphics[width=245pt]{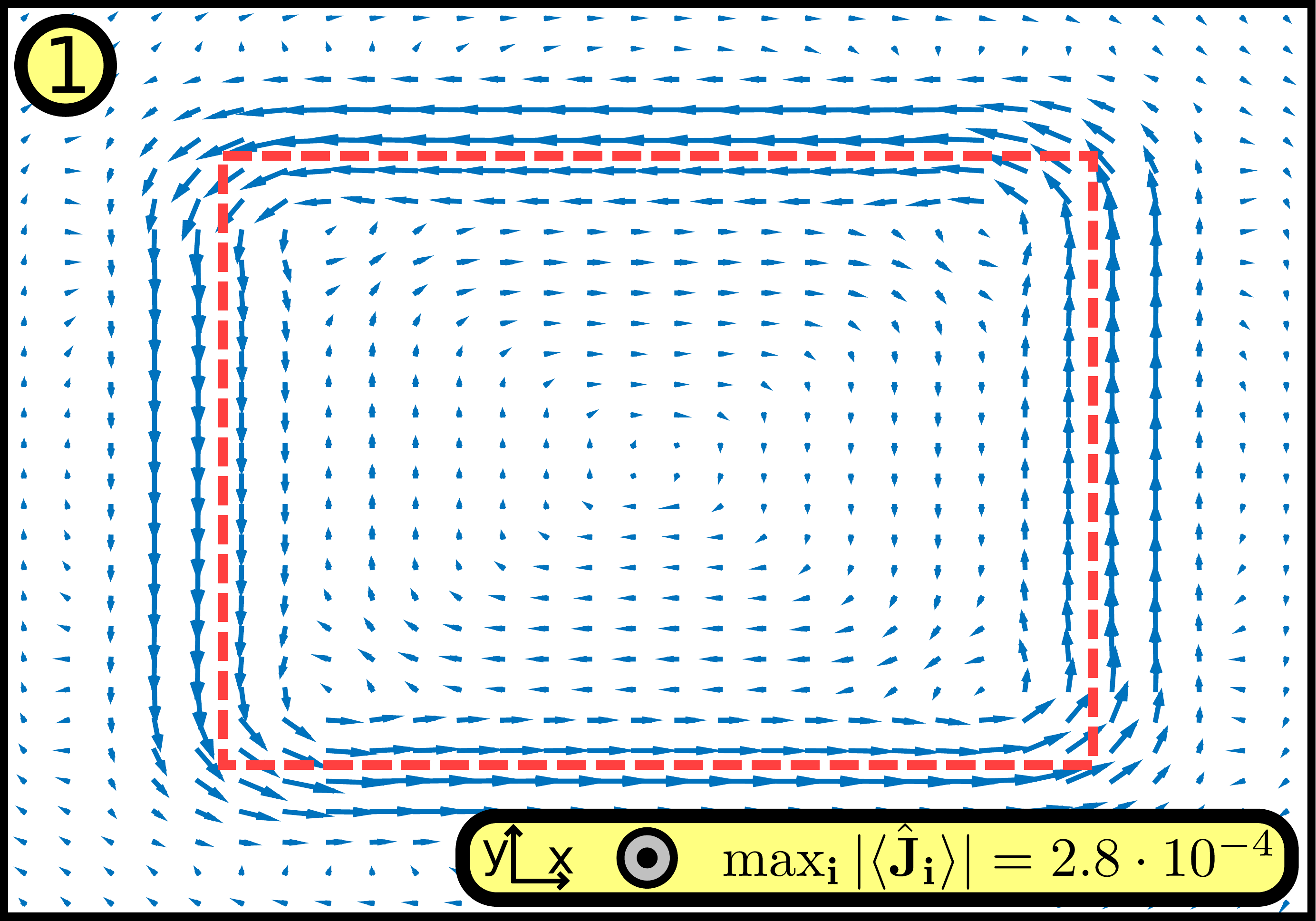}
\caption{(Color online.) Edge currents in the topologically trivial phase of a ferromagentic block magnetized along the direction $\hat{\mathbf{n}} = \hat{\mathbf{z}}$, as indicated by grey arrows.
The current density is proportional to the arrow length, with the longest arrow corresponding to a current density of $2.8\cdot 10^{-4}$.
See also (1) in Fig.~\ref{Figure:D_Jmax_block_perp}.
}
\label{Figure:J_73_block_perp}
\end{figure}

\begin{figure}
\includegraphics[width=245pt]{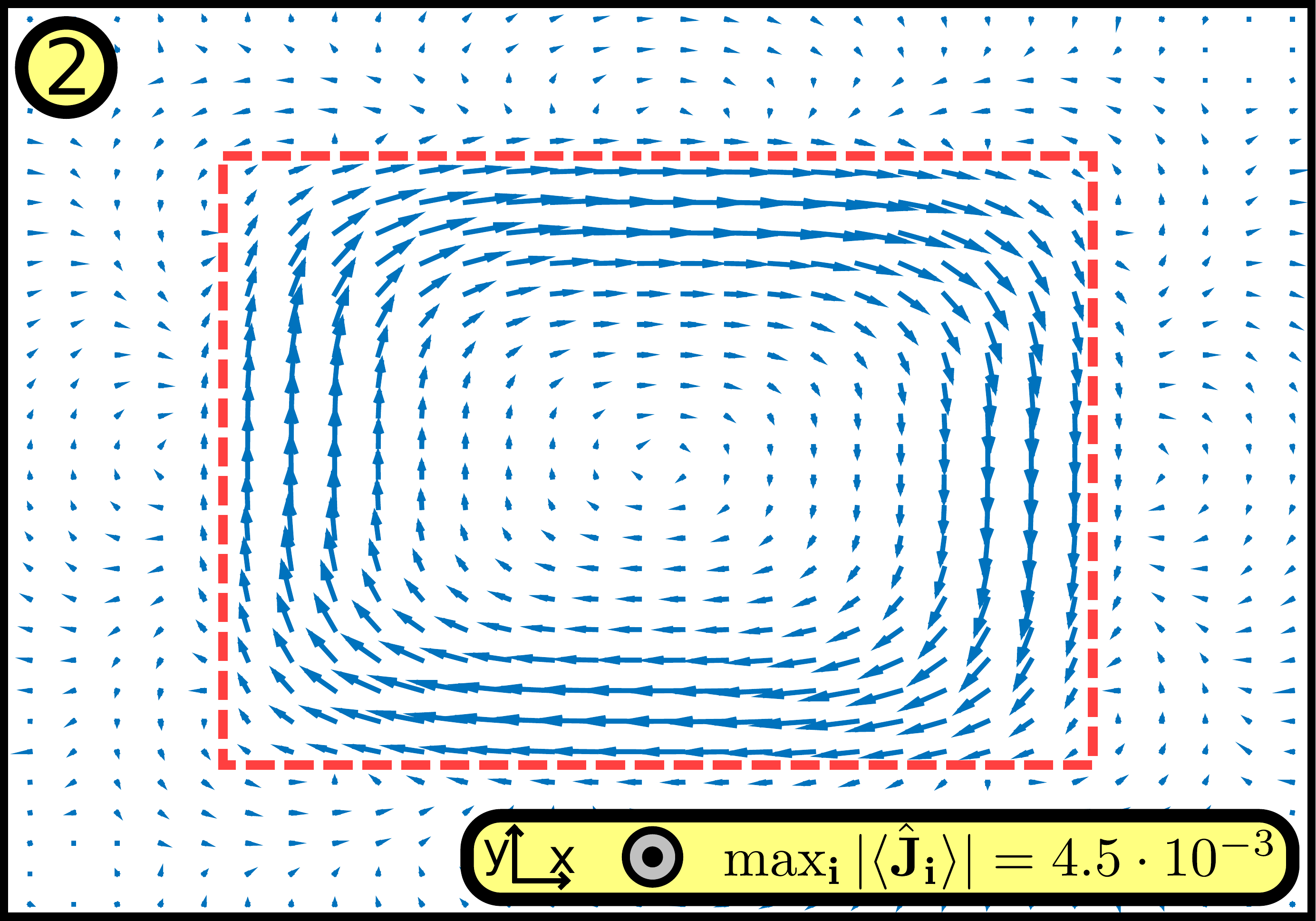}
\caption{(Color online.) Edge currents in the topologically nontrivial phase of a ferromagnetic block magnetized along the direction $\hat{\mathbf{n}} = \hat{\mathbf{z}}$.
See also (2) in Fig.~\ref{Figure:D_Jmax_block_perp}.
}
\label{Figure:J_80_block_perp}
\end{figure}

\begin{figure}
\includegraphics[width=245pt]{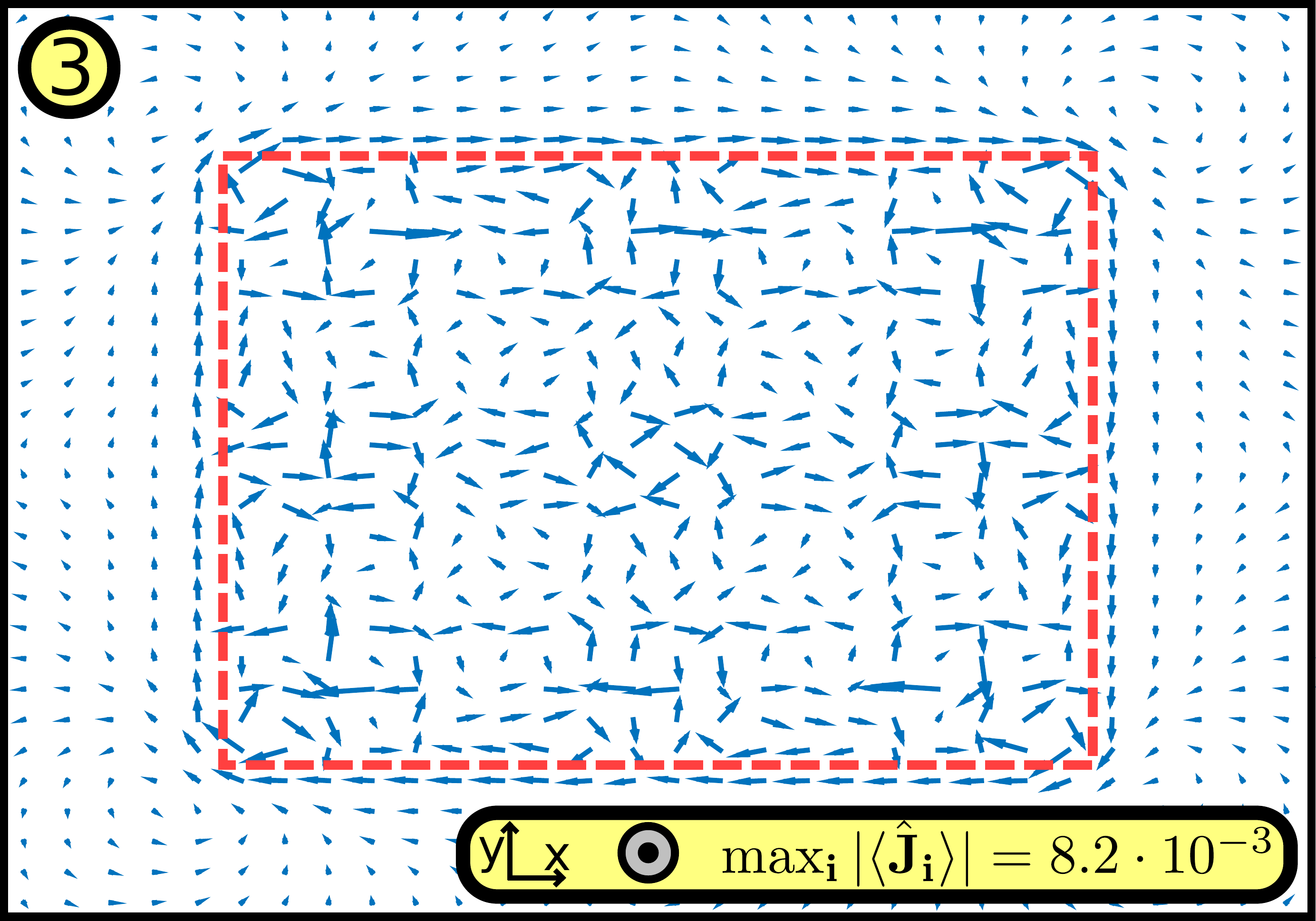}
\caption{(Color online.) Currents at a current density peak in the collapsed gap state of a ferromagnetic block magnetized along the $\hat{\mathbf{n}} = \hat{\mathbf{z}}$.
See also (3) in Fig.~\ref{Figure:D_Jmax_block_perp}.
}
\label{Figure:J_111_block_perp}
\end{figure}

We begin our investigation with a quadratic ferromagnetic block, with the magnetic moments perpendicular to the surface ($\hat{\mathbf{n}} = \hat{\mathbf{z}}$).
In Fig.~\ref{Figure:D_Jmax_block_perp} the maximum value of the current density and the size of the order parameter at the center of the ferromagnetic block is plotted as a function of the Zeeman term, which quantify the strength of the magnetic impurity moments.
The condition for being in the topologically nontrivial phase, given in Eq.~\eqref{Equation:Topological_phase_transition_condition}, is fulfilled in the shaded regions.
It is clear that the maximum current density is comparatively small in the trivial phase, raises up inside the topologically nontrivial phase, and then falls off as superconductivity is destroyed in the ferromagnetic block.
We also plot the spatial distribution of the persistent currents in Figs.~\ref{Figure:J_73_block_perp}-\ref{Figure:J_111_block_perp} for the three points marked 1-3 in Fig.~\ref{Figure:D_Jmax_block_perp}.
For small Zeeman splitting the current flows anti-clockwise along the edge of the block, in agreement with a recent Ginzburg-Landau calculation for a circular disk geometry.\cite{arXiv:1505.01672}
Next, in Fig.~\ref{Figure:J_80_block_perp}, the persistent current in the nontrivial phase is seen to not only be much larger, but it also flows in the opposite direction to that in the trivial phase.
We also note that this current is located mainly inside the ferromagnetic region.
Below we show that these currents are related to the appearance of edge states, but surprisingly propagates in the  direction opposite to what is expected from the edge state spectrum.
However, before we move on to a more detailed study of the currents in the nontrivial phase, which is the main focus of this work, we also say a few words about the currents in the third region.


We see that the maximum current density remains fairly large even after superconductivity is destroyed in the ferromagnetic block region, and these currents are also located around the edge of the block.
Strong edge currents therefore remain even after the superconducting gap has been destroyed, even though it is not possible to motivate their presence using topological arguments for the block region.
A few spikes also appear in the maximum current density, and these are due to the formation of vortex-like currents inside the block region, as illustrated in Fig.~\ref{Figure:J_111_block_perp}.
The exact occurrence of these peaks as a function of the magnetization is dependent on the size of the block and these current patterns result from the interplay of the Zeeman term, Rashba spin-orbit interaction, and boundary conditions.
A complete study of the currents in this regime is beyond the scope of this work, and we here only mention these effects to clarify the properties of the non-zero current in the third region.

\subsubsection{Reversed persistent current}
\label{Subsubsection:Reversed_persistent_current}
\begin{figure}
\includegraphics[width=245pt]{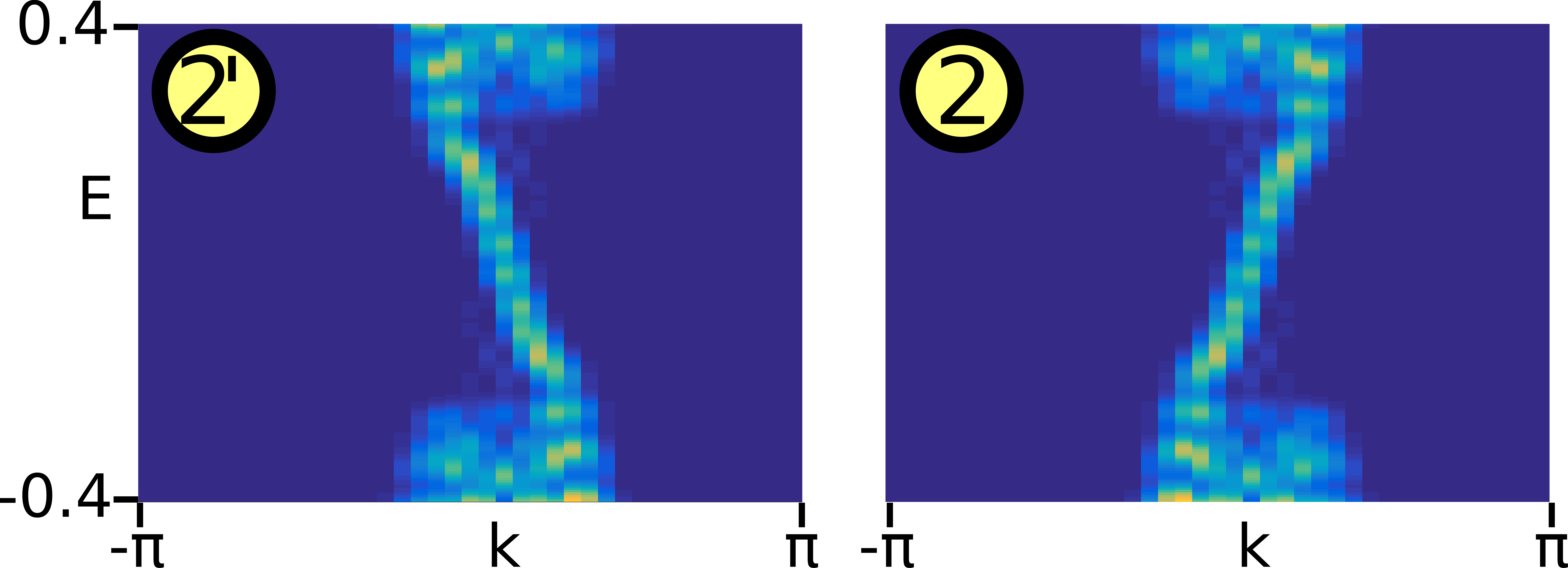}
\caption{(Color online.) Spectral function along the lower edge of a ferromagnetic block magnetized along the direction $\hat{\mathbf{n}} = \hat{\mathbf{z}}$.
Only contributions from states classified as edge states are included.
Edge state on the left and right corresponds to the topologically nontrivial phases at negative and positive Zeeman term, respectively, see (2') and (2) in Fig.~\ref{Figure:D_Jmax_block_perp}.
The opposite dispersions reflect the opposite signs of the Chern number in the two phases.}
\label{Figure:Spectral_function}
\end{figure}

Having covered the basic behavior of the currents, we now investigate in detail the current in the nontrivial phase and relate it to the topologically protected edge states.
In Fig.~\ref{Figure:Spectral_function} we plot the contribution to the spectral function from states classified as edge states for a cut along the lower edge of the ferromagnetic block.
The left and the right figure corresponds to the nontrivial phase that occurs for negative and positive Zeeman term, respectively.
It is clear that the edge states have opposite dispersion, in agreement with the two phases having opposite Chern numbers.
We now note that the slope of the edge state in panel (2) of Fig.~\ref{Figure:Spectral_function} implies that the edge state has a propagation direction to the right, since $v \sim dE/dk$.
However, this is in contradiction to the direction of the persistent current plotted in Fig.~ \ref{Figure:J_80_block_perp}, which flows to the left along the lower edge.

\begin{figure}
\includegraphics[width=245pt]{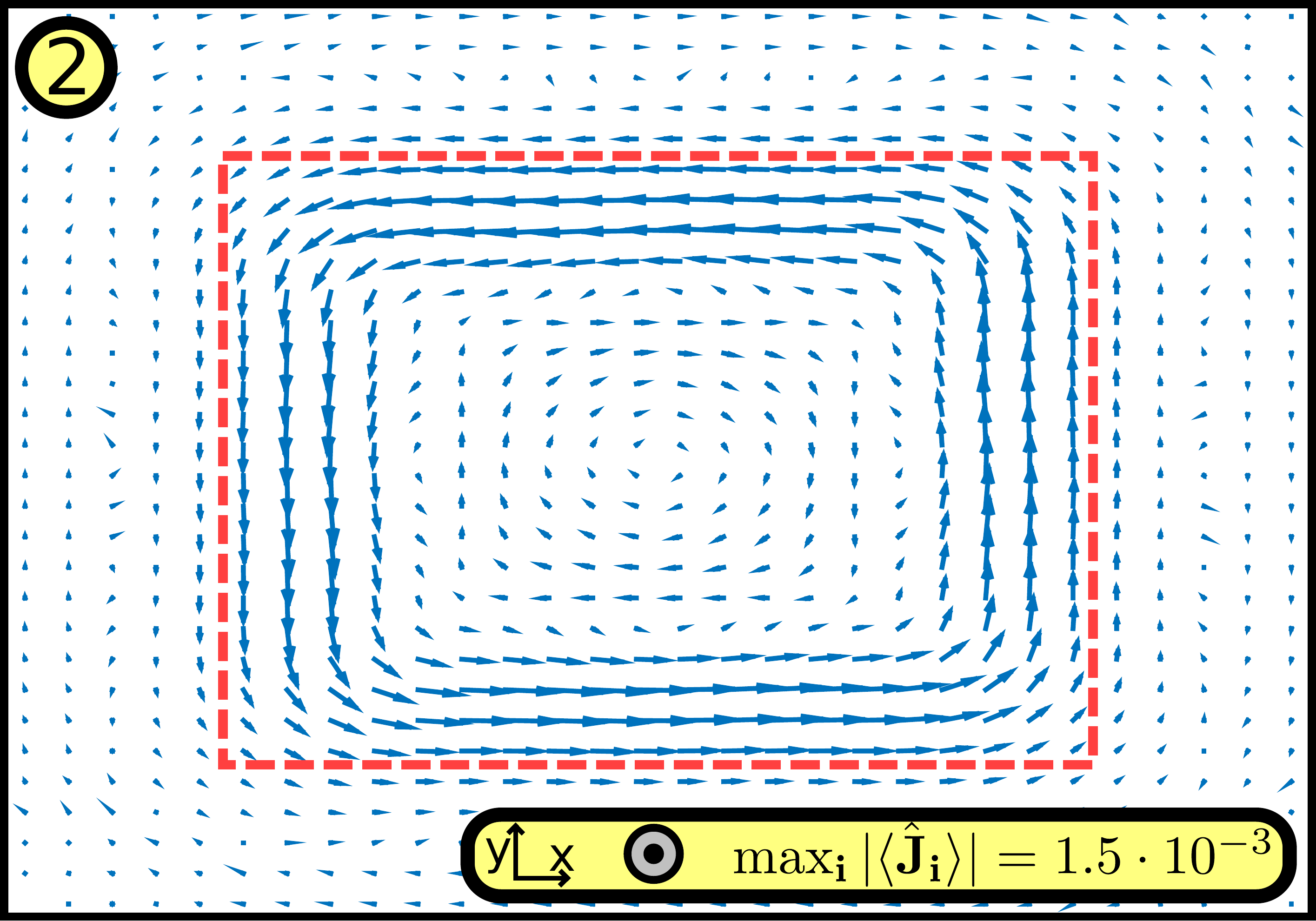}
\caption{(Color online.) Same as in Fig.~\ref{Figure:J_80_block_perp}, but when only contributions from eigenstates with energy in the interval $-0.1 < E < 0$ have been included.}
\label{Figure:J_edge_contribution}
\end{figure}

The apparent contradiction is resolved once the contribution to the current from the edge states is singled out.
In Fig.~\ref{Figure:J_edge_contribution} we plot the contribution to the current from eigenstates with energy in the narrow interval $-0.1 < E < 0$ around the Fermi level, and the current is seen to flow in opposite direction to the total current in Fig.~\ref{Figure:J_80_block_perp}.
The somewhat unexpected conclusion is therefore that the persistent currents flows in opposite direction to the current carried by the topologically protected gap-crossing edge states.
This have important experimental consequences as the measured current direction will depend on the physical probe used.
If for example a SQUID is used to measure the magnetic field generated by the total persistent current, the current will be found to flow in opposite direction compared to the dispersion of the edge states.
On the other hand, any transport measurement relying on excitations close to the Fermi level will find a propagation direction in agreement with the edge state dispersion.

The reason for the reversed flow of the persistent currents compared to the edge state currents can be understood by considering why an ordinary bulk is free from persistent currents.
This is not due to zero contribution to the current from individual charge carriers, but rather because the sum of all such contributions cancel each other out.
The total current carried by the system is obtained by integrating the product of the occupied charge carriers charge and velocity: $J \sim \int_{E(\mathbf{k})<k_F} e\nabla_{\mathbf{k}}Ed\mathbf{k} = e\int_{E(\mathbf{k})<k_F}\nabla_{\mathbf{k}}Ed\mathbf{k} = 0$.
Here the first step follows because the charge of each quasiparticle state is the same, while the second step follows because the spectrum is continuous and periodic in $\mathbf{k}$.
Neither of these assumptions are true at the edge of a topological superconductor.
First of all, a branch in the spectral function which crosses the Fermi level an odd number of times cannot be described by a simultaneously continuous and periodic spectrum.
Second, the quasiparticles in a superconductor are mixtures of electrons and holes and do not all carry the same charge.
Considering the persistent current carried by the edge states, we now note that only half of the branch corresponds to occupied states.
Further, the occupied edge states are strong mixtures of electrons and holes because they are close to the Fermi level.
The total current carried by this band is therefore much smaller than what would be expected if the band corresponded to a fully occupied electron band.
It is this lack of occupied electron states associated with the gap-crossing band which fails to cancel the counter-propagating currents coming from the low energy states.
This phenomenon of total currents adding up from contributions of quasiparticle currents of states "inside the gap" and deep states that contribute to overall condensate flow is known in unconventional superconductors and superfluids.\cite{PhysRevB.84.214509, PhysRevB.88.184506}

\subsubsection{Spin-polarized current}
\begin{figure*}
\includegraphics[width=245pt]{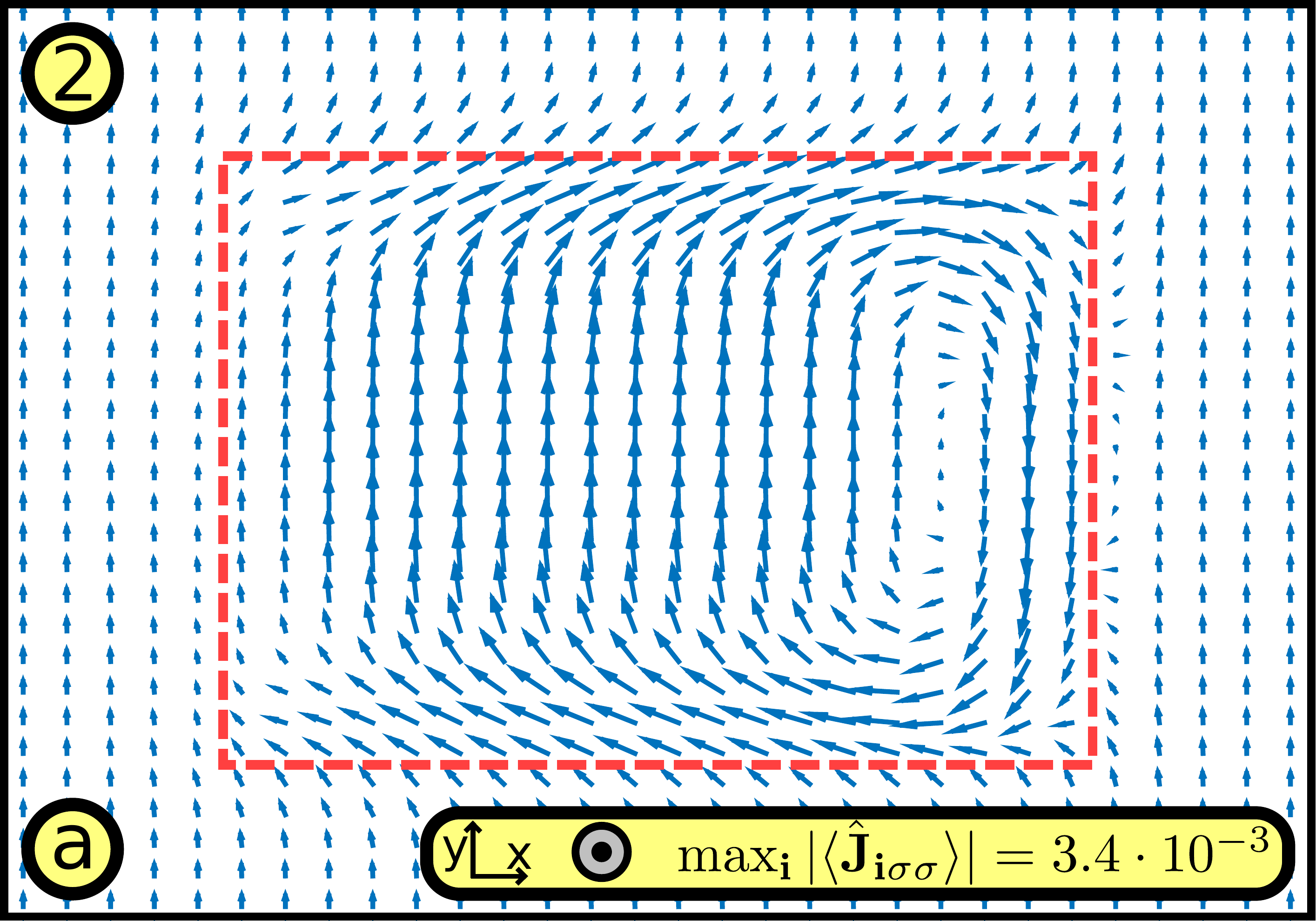}
\includegraphics[width=245pt]{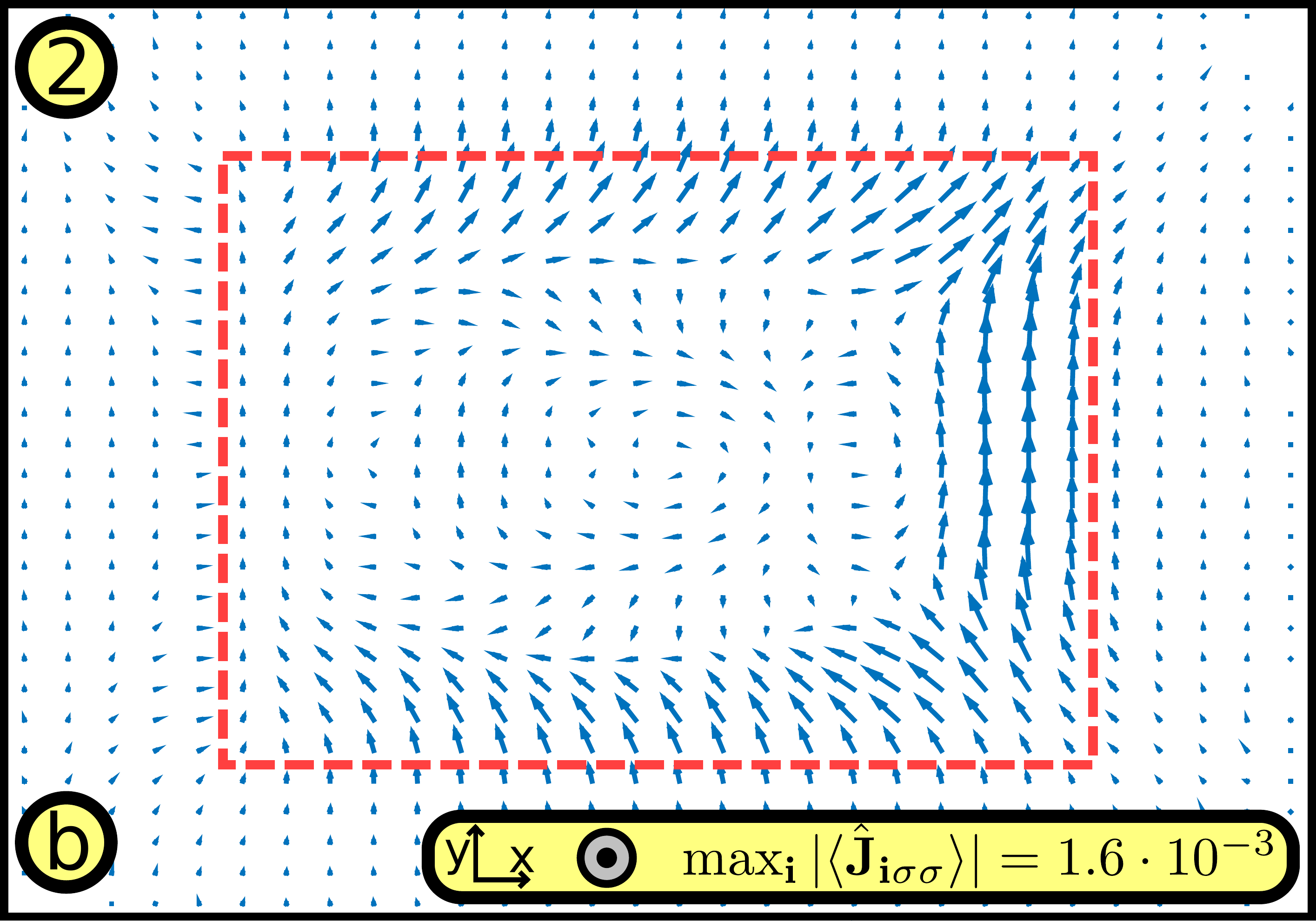}
\caption{(Color online.) Spin-polarized current for spins along the positive $x$-axis for a ferromagnetic block magnetized along the direction $\hat{\mathbf{n}} = \hat{\mathbf{z}}$.
(a): Total spin-polarized current, (b): only contributions from states with energy in the interval $-0.1 < E < 0$.
The current pattern is related to the spin-polarized current for $y$-up, $x$-down, and $y$-down spins by a spatial rotation of $\pi/2$, $\pi$, and $3\pi/2$, respectively. 
}
\label{Figure:J_spin_polarized}
\end{figure*}

Next we study the spin-polarization of the currents.
In particular, the $x$ and $y$ spin-polarization axes are of interest because the Rashba spin-orbit interaction couples these axes differently to the momentum.
In Fig.~\ref{Figure:J_spin_polarized} we plot the total spin-polarized current as well as that carried by eigenstates in the energy interval $-0.1 < E < 0$ for spins polarized along the $x$-axis.
The spin-polarized currents for $y$-up, $x$-down, and $y$-down spins are identical to these under a spatial rotation of $\pi/2$, $\pi$, and $3\pi/2$, respectively.
It is clear from Fig.~\ref{Figure:J_spin_polarized}(a) that strong spin-polarized currents flow through the block, but the total current in the bulk cancels, since the spins polarized along the negative $x$-axis flow in opposite direction.
However, in the edge regions the two currents flow in the same direction and, together with the 'spin-flipping' currents, these gives rise to the total edge currents in Fig.~\ref{Figure:J_80_block_perp}.

We also note the partial depletion of $x$-up spin currents towards the right end of the block in Fig.~\ref{Figure:J_spin_polarized}(a).
The picture is radically different in Fig.~\ref{Figure:J_spin_polarized}(b), where only contributions to the currents from states close to the Fermi level are included.
Here, the $x$-up spins are located on the right side, and with a similar argument to that in Sec.~\ref{Subsubsection:Reversed_persistent_current}, we can understand the depletion of the total $x$-up spin current in Fig.~\ref{Figure:J_spin_polarized}(a) as a partial, but incomplete, cancellation by the states close to the Fermi level.

It is clear from both the total current in the bulk region and the edge currents in Fig.~\ref{Figure:J_spin_polarized}(b) that $x$-up spins favor motion along the positive $y$-axis.
This can be understood as a consequence of the presence of Rashba spin-orbit interaction.
The Rashba spin-orbit interaction couples momentum and spin through $\mathbf{k}\times\boldsymbol{\sigma}$, which leads to spin-momentum locking for the currents.\cite{arXiv:1505.01672}
This spin-momentum locking and localization of different spins on different edges are, as we will discuss next, also important for understanding 1D wire systems.

\subsection{Wire with magnetic moments perpendicular to surface}
\label{Subsection:Wire_with_Zeeman_term_perpendicular_to_surface}
We now move on to an investigation of a 1D wire with the magnetic moments still perpendicular to the surface, which corresponds to the 1D model in Eq.~\eqref{Equation:Bulk_Hamiltonian_1D}.
As explained in Sec.~\ref{Section:Dimensional_reduction}, we no longer have a simple condition, such as Eq.~\eqref{Equation:Topological_phase_transition_condition}, to rely on to identify the topological phase transition, because the system is embedded in a 2D surface rather than being truly 1D.
However, we find that this present little problem since we have several other means to identify the nontrivial phase.

\subsubsection{Partial gap collapse and topological phase transition}
\begin{figure}
\includegraphics[width=245pt]{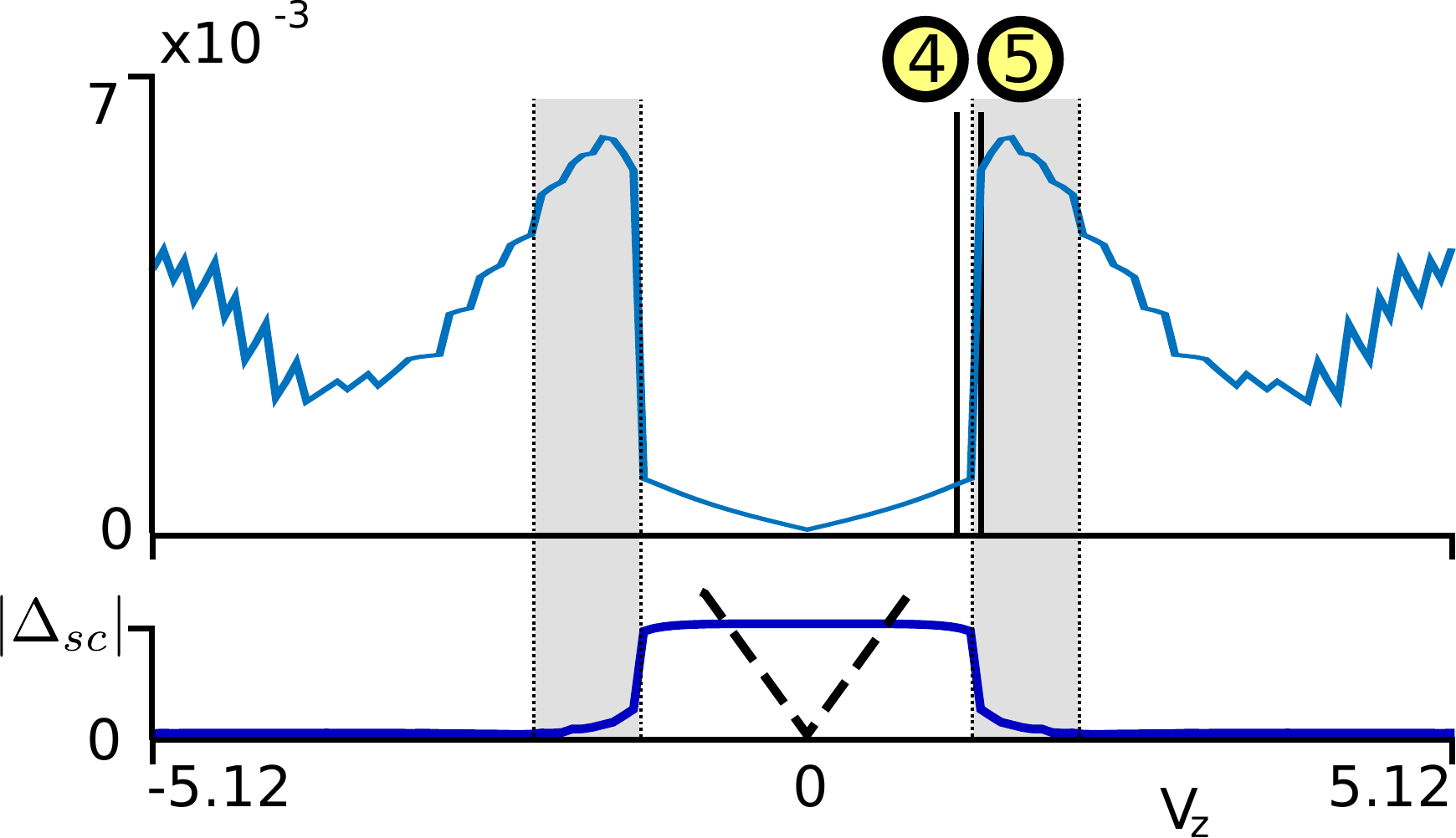}
\caption{(Color online.) Similar to Fig.~\ref{Figure:D_Jmax_block_perp}, but for a 1D wire along the $x$-axis and magnetic moments along the direction $\hat{\mathbf{n}} = \hat{\mathbf{z}}$.
The topologically nontrivial phase is identified using the discontinuous onset of strong edge currents as well as the appearance of Majorana fermions.}
\label{Figure:D_Jmax_w2_perp}
\end{figure}

In Fig.~\ref{Figure:D_Jmax_w2_perp} we plot the maximum current density and the value of the order parameter at the midpoint of a wire.
First of all, we note that compared to Fig.~\ref{Figure:D_Jmax_block_perp} the horizontal axis has been rescaled and, compared to the block, a roughly twice as strong magnetization is required to suppress superconductivity on the wires sites.
A similar calculation for a point impurity reveals that the magnetization has to be about four times as large as as that of the block geometry in order to destroy superconductivity.
This can be understood from the fact that each edge site in a block is connected to a single nonmagnetic superconducting site, while each site in the wire and for the single impurity is connected to two and four such superconducting sites, respectively.

Next, we note that the partial collapse of the order parameter in itself does not imply a topological phase transition, even though we numerically always see these occurring at the same time.
There are, however, good reasons for why these can be expected to occur at the same time.
First of all, the topological phase transition occurs when $\Delta$ and $V_z$ are related according to an expression similar to Eq.~\eqref{Equation:Topological_phase_transition_condition}, even though we can no longer rely on the last equality sign or the exact values of $t$ and $\mu$.
The inequality is clearly valid when making the Zeeman term large enough, but also if the order parameter is small enough.
A sudden drop in the order parameter will therefore likely induce the topological phase transition.
Thus, assuming no a priori knowledge of the value of $\Delta$ at which the transition occurs, the likelihood of the transition to occur is proportional to the size of the drop in the order parameter.

An even stronger argument for why the two phenomena should occur in tandem can be made if we instead consider the drop in the order parameter to be induced by the topological phase transition.
This relies on the fact that a topological phase transition is fundamentally related to the interchange of energy levels across the Fermi level, also known as a band inversion.
Because the order parameter is calculated using eigenstates below the Fermi level, the value of the order parameter can be expected to experience discontinuous changes each time a pair of eigenstates are interchanged across the Fermi level.
Moreover, because the size of the order parameter in turn affects the eigenstates, this has enhancing feedback effects, which are only captured in fully self-consistent calculations such as those performed here.
Discontinuous changes in the order parameter and the topological phase transition are therefore intimately tied to each other. 
We note that this is closely related to the physics of single magnetic impurities.
The associated Yu-Shiba-Rusinov states, which in non-self-consistent calculations cross the Fermi level continuously, cross discontinuously in self-consistent models because of the feedback effect of the occupation numbers of these states on the order parameter.\cite{arXiv:1505.01672, RevModPhys.78.373}

\subsubsection{Signatures of nontrivial topology}
\begin{figure}
\includegraphics[width=245pt]{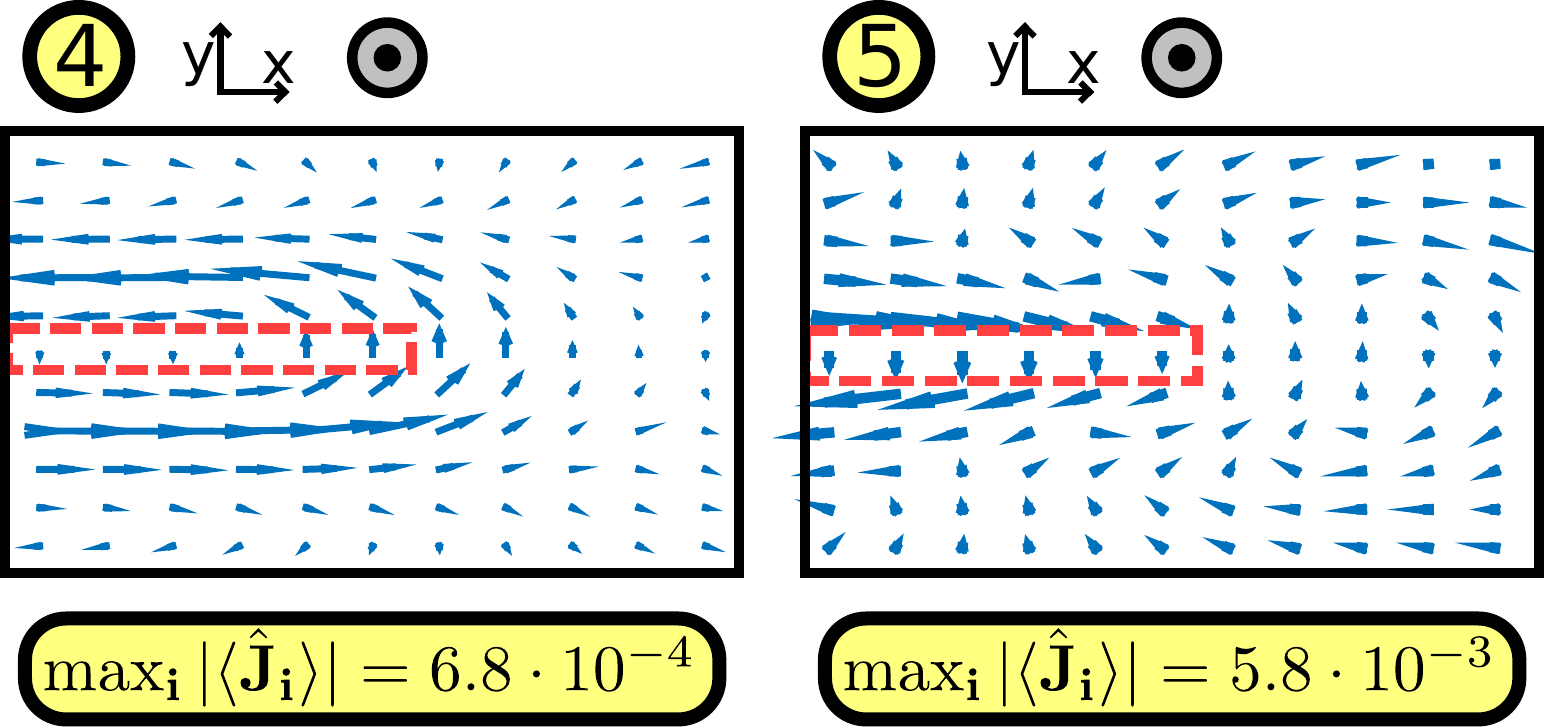}
\caption{(Color online.) Persistent currents around ferromagnetic wire with magnetic moments along the direction $\hat{\mathbf{n}} = \hat{\mathbf{z}}$.
Left: Trivial phase. Right: Nontrivial phase.
See also (4) and (5) in Fig.~\ref{Figure:D_Jmax_w2_perp}.
}
\label{Figure:J_w2_perp}
\end{figure}

\begin{figure}
\includegraphics[width=245pt]{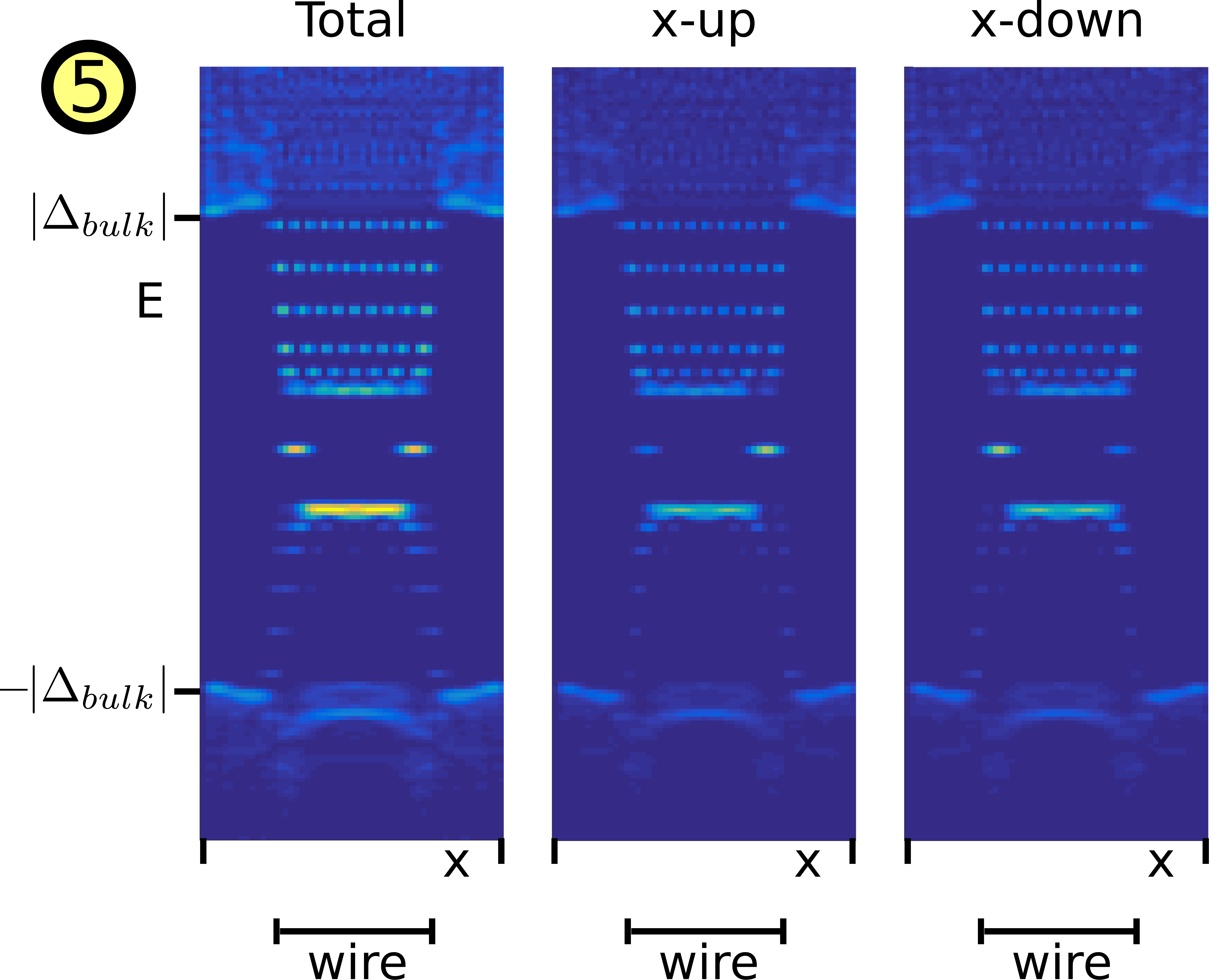}
\caption{(Color online.) LDOS as a function of position along a 1D wire when magnetic moments are perpendicular to the surface ($\hat{\mathbf{n}} = \hat{\mathbf{z}}$).
Both total and spin-polarized LDOS are shown, using the $x$-axis as spin-polarization axis.
The extent of the wire is indicated at the bottom.
One Majorana fermion is clearly seen at each end of the wire.
}
\label{Figure:LDOS}
\end{figure}

So far we have discussed the relation between the collapse of the gap and the topological phase transition, arguing that there are strong reasons to expect them to occur at the same time.
We have also mentioned that they always occur at the same time in our calculations, but so far not said what signatures we have used to identify the topological phase transition.
The first signature is the strong peak in the maximum current density in Fig.~\ref{Figure:D_Jmax_w2_perp}.
Just like for the square block, the onset of the peak is also associated with a reversal of the direction of the persistent current, as can be seen in Fig.~\ref{Figure:J_w2_perp}.
Most important though, this is also the point where Majorana fermions appear at the end points of the wire, as shown in Fig.~\ref{Figure:LDOS}. The Majorana fermion nature of these zero-energy states have been confirmed by checking that they appear as locally non-degenerate Bogoliubov-de Gennes quasiparticles.

\subsubsection{Majorana spin-polarization and block edge currents}
An interesting feature of the Majorana fermions seen in Fig.~\ref{Figure:LDOS} are their spin-polarizations.
The Majorana fermions at the right and left end of the wire are polarized along the $x$-up and $x$-down direction, respectively.
That Majorana fermions have spin-polarization has already been predicted,\cite{PhysRevLett.108.096802} but we are here able to understand it in relation to the spin-polarized currents around a 2D ferromagnetic block.
In Fig.~\ref{Figure:J_spin_polarized}(b) it was shown that the contribution to the $x$-up polarized current coming from states close to the Fermi level is located on the right edge.
This is in complete agreement with the $x$-up polarized Majorana fermion localized on the right end of the wire.
This demonstrates the intimate connection between the topologically protected edge states of a 2D and 1D system.

\begin{figure}
\includegraphics[width=245pt]{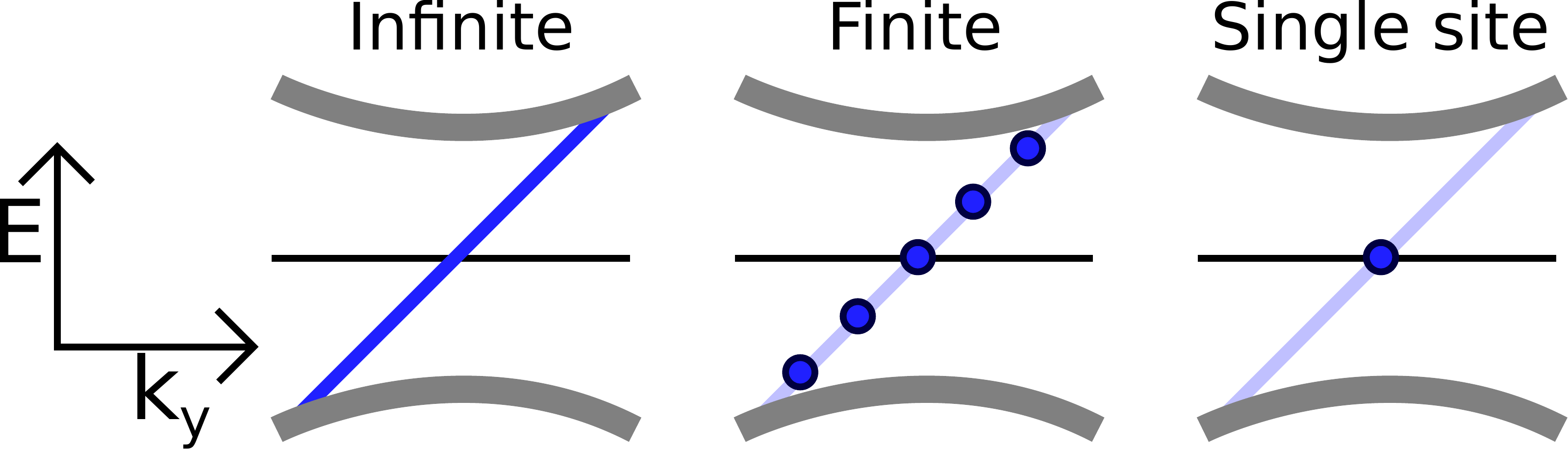}
\caption{(Color online.) Schematic figures showing the edge states along the rightmost edge of a ferromagnetic strip with varying length.
If the edge is made infinitely long the spectrum is continuous, but for any finite length, the spectrum becomes discrete.
For a 1D wire placed along the $x$-axis, the Majorana fermion at its end point can be understood as the single state that remains when the length of the $y$-edge becomes a single site.
}
\label{Figure:Edge_states_dimensional_reduction}
\end{figure}

The connection between edge states and Majorana fermions can be made even clearer by considering the edge states plotted in Fig.~\ref{Figure:Spectral_function}.
For simplicity of argument, we first consider an infinite ferromagnetic strip rather than a square block and let the infinite dimension be along the $y$-direction.
On each of the two edges one spectral branch cuts the Fermi level, one of which is schematically depicted in the leftmost picture of Fig.~\ref{Figure:Edge_states_dimensional_reduction}.
Next, consider the $y$-dimension wrapped around a cylinder, such that it is still translationally invariant but with finite length.
This results in a discretization of the spectral function as shown in the middle picture of Fig.~\ref{Figure:Edge_states_dimensional_reduction}.
By finally making the cylinder small enough such that the $y$-direction consists of a single site, a 1D wire with a single Majorana fermion at each edge results, as depicted to the right in Fig.~\ref{Figure:Edge_states_dimensional_reduction}.
This procedure closely resembles the dimensional reduction performed in Sec.~\ref{Section:Dimensional_reduction}, and explicitly shows the close relation between Majorana fermions in 1D and edge states in 2D.
However, this dimensional reduction does not, just like that performed in Sec.~\ref{Section:Dimensional_reduction}, correspond to the actual physical situation.
This is important to recognize because it might otherwise give the false impression that it is enough to discretize the edge spectrum to obtain a localized Majorana fermion.
After all, the state at $E = 0$ in the middle panel of Fig.~\ref{Figure:Edge_states_dimensional_reduction} is also a Majorana fermion.

The reason why the relation between the 2D block and the 1D wire is slightly more complicated is that the two $x$-edges, which can be treated completely independent for the infinite strip, is joined by two $y$-edges.
The spectral branches of the different edges therefore join into one single branch with eigenstates delocalized over the whole boundary.
In particular, the Majorana fermions that otherwise would have existed at the right and left edge overlap and hybridize.
It is only once the $y$-dimension becomes small enough, such that the two $y$-edges start hybridizing and thereby gaps out the spectrum along these edges, that the two $x$-edges becomes independent of each other.
It is therefore clear why it is not enough to have a discrete spectrum, such as for a block with finite edge length, to obtain localized Majorana fermions, and why a (quasi-)1D wire is required in spite of the close connection between the two systems.

\subsection{Block with magnetic moments parallel to surface}
\label{Subsection:Block_with_Zeeman_term_parallel_to_surface}
Having investigated magnetic moments perpendicular to the surface, we now turn to a ferromagnetic block with the magnetic moments pointing along the $x$-direction ($\hat{\mathbf{n}} = \hat{\mathbf{x}}$).
This system does not have a topologically nontrivial phase, but is of interest because of its relation to the corresponding 1D wires.

\begin{figure}
\includegraphics[width=245pt]{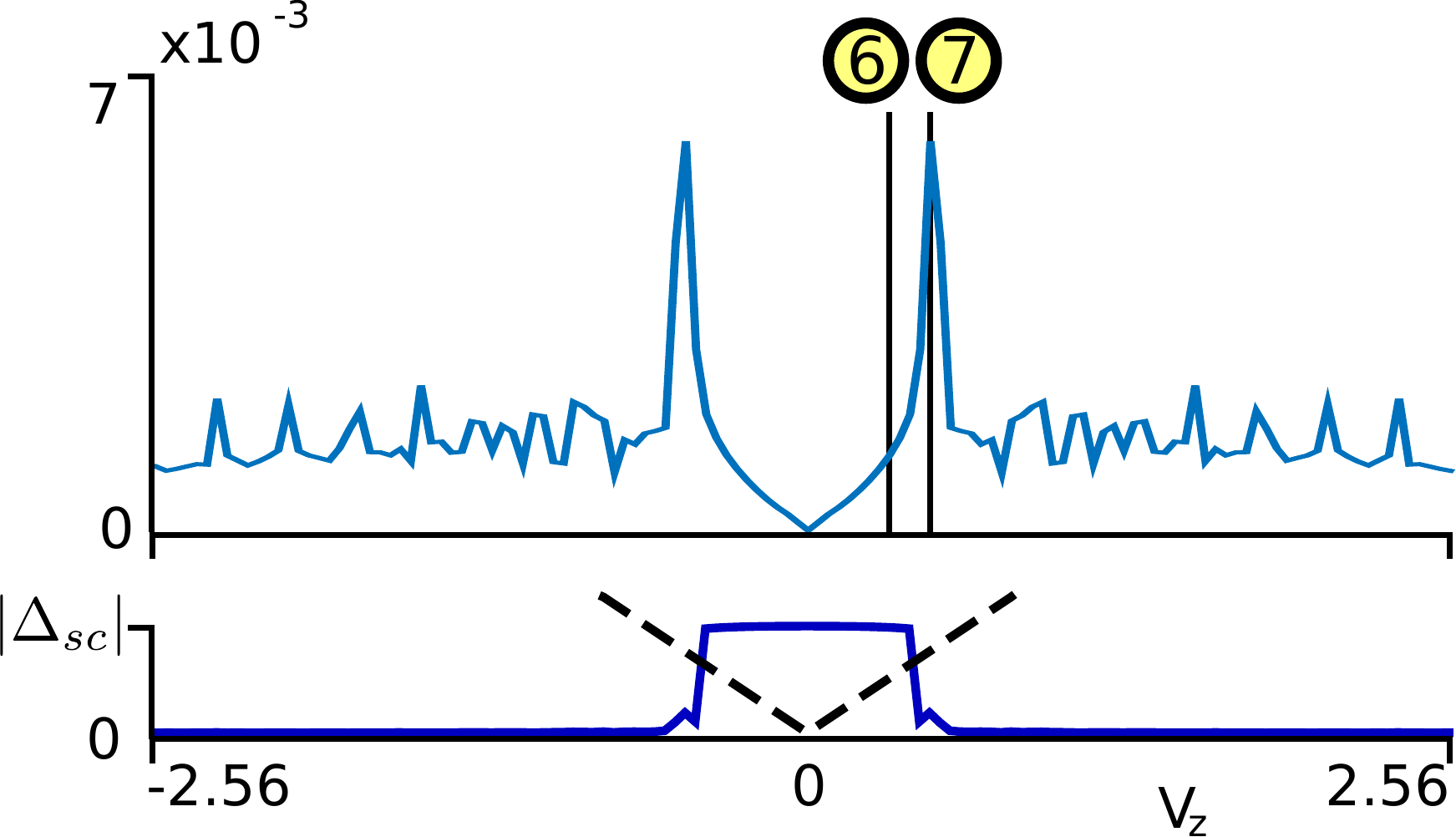}
\caption{(Color online.) Same as in Fig.~\ref{Figure:D_Jmax_block_perp}, but for magnetic moments along the direction $\hat{\mathbf{n}} = \hat{\mathbf{x}}$.
}
\label{Figure:D_Jmax_block_para}
\end{figure}

\begin{figure}
\includegraphics[width=245pt]{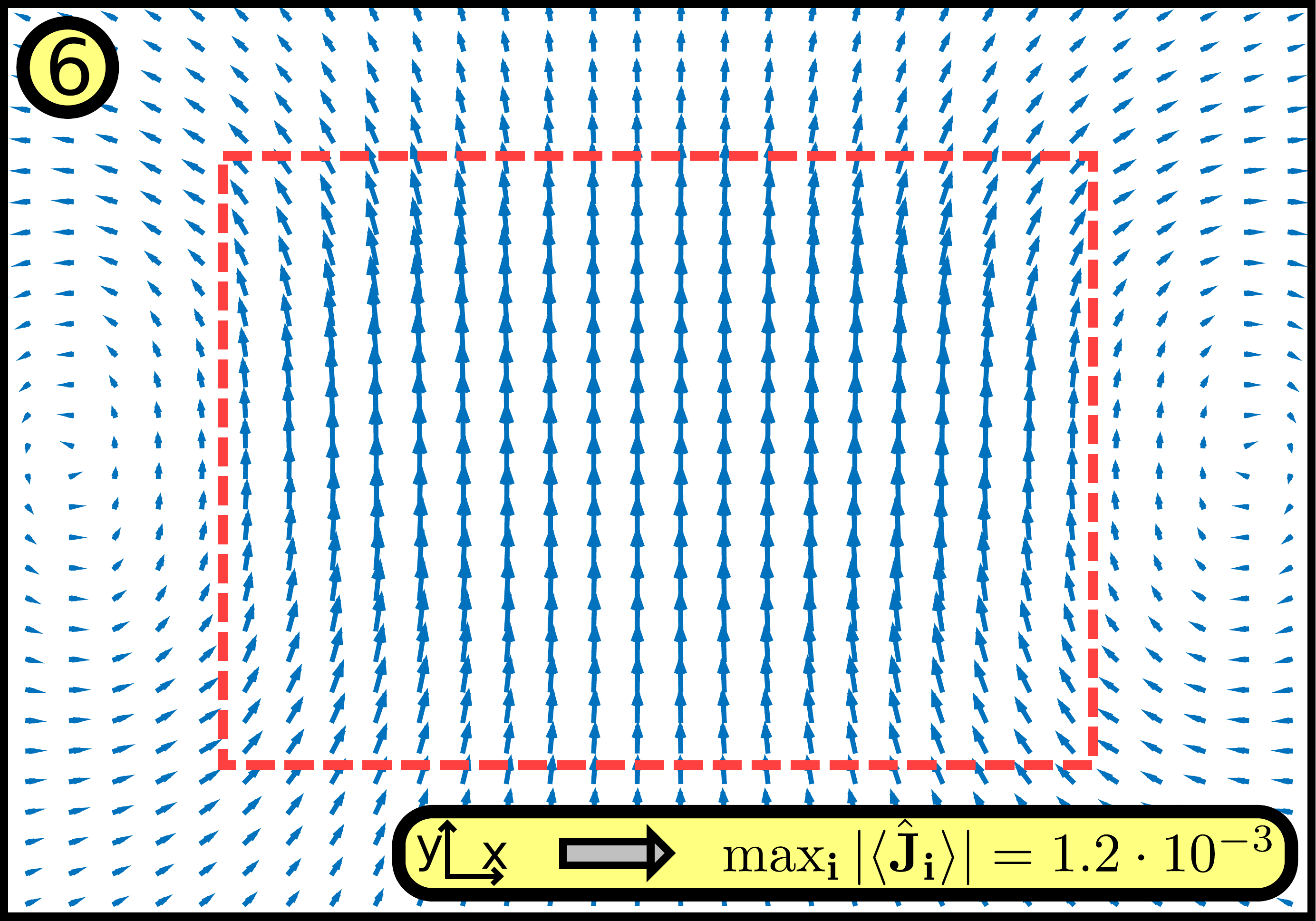}
\caption{(Color online.) Persistent currents before the partial collapse of the superconducting gap for a ferromagnetic block magnetized along the direction $\hat{\mathbf{n}} = \hat{\mathbf{x}}$.
See also (6) in Fig.~\ref{Figure:D_Jmax_block_para}.
}
\label{Figure:J_73_block_para}
\end{figure}

In Fig.~\ref{Figure:D_Jmax_block_para} we plot the order parameter and maximum current density.
Interestingly, a strong peak is observed also in this case, even though there is no topological phase transition.
However, the build up of the peak is not discontinuous as in the previous cases.
Instead, the maximum current density is sizable already before superconductivity starts to collapse in the ferromagnetic block.
In fact, it is not only the maximum current density that is large, but also the total persistent current.
This is clear from Fig.~\ref{Figure:J_73_block_para}, where it is seen that the current flows through the whole sample, rather than being limited to the boundary of the ferromagnetic block.
A similar behavior is predicted using a Ginzburg-Landau model in Appendix \ref{Appendix:Ginzburg_Landau}.

\begin{figure}
\includegraphics[width=245pt]{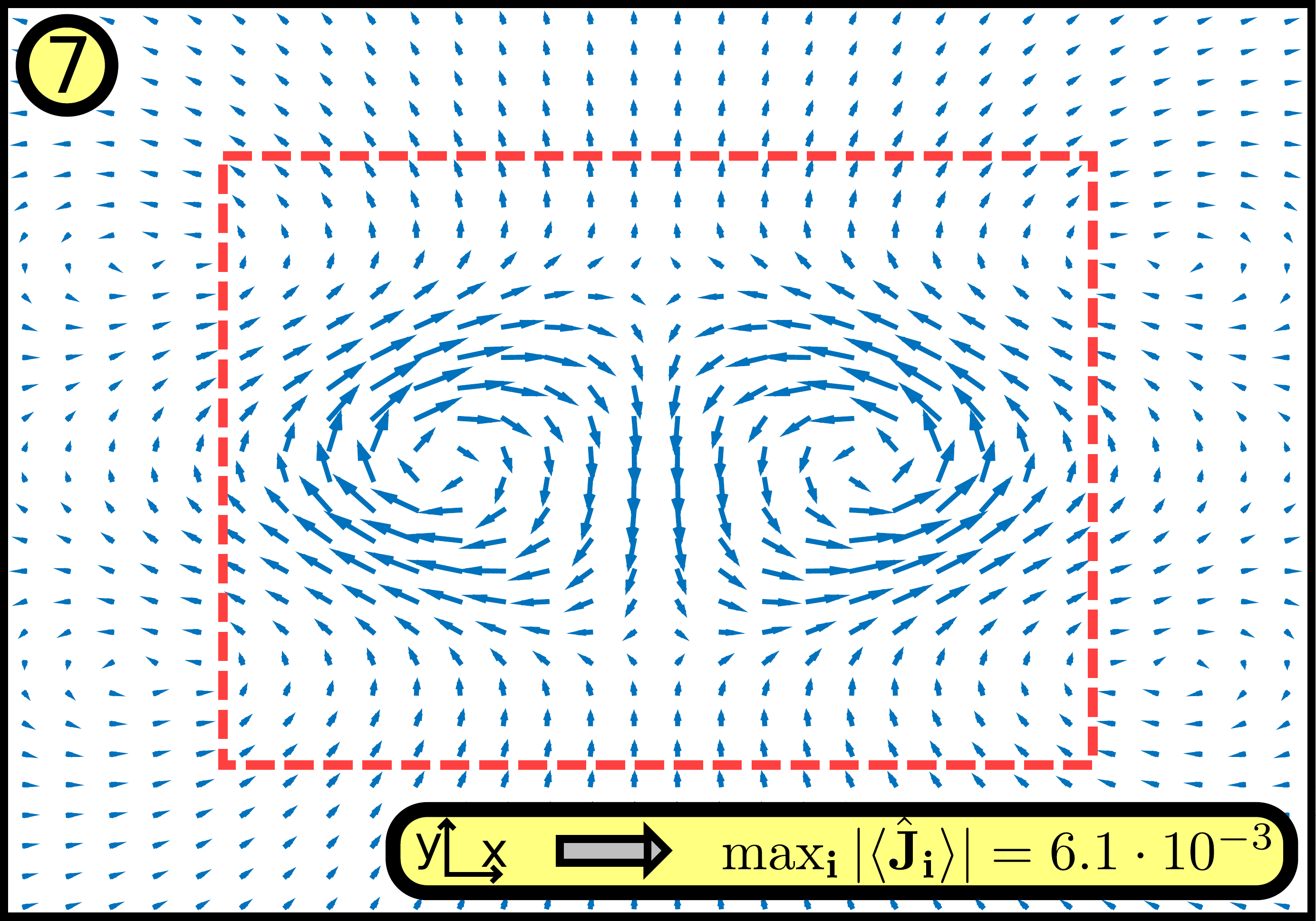}
\caption{(Color online.) Persistent current pattern at maximum current density peak at (7) in Fig.~\ref{Figure:D_Jmax_block_para}.
Vortex-like solutions continue to appear in the ferromagnetic region also for somewhat larger Zeeman terms, until the current eventually becomes localized on the left and right edges.
}
\label{Figure:J_77_block_para}
\end{figure}

The current flowing through the bulk region of the ferromagnetic block can be understood as a consequence of $x$-up and $x$-down spins moving in opposite direction because of spin-orbit interaction, similar to what was seen in Fig.~\ref{Figure:J_spin_polarized}.
The addition of a Zeeman term in the $x$-direction creates an imbalance between the two spin-species and thereby also a total current through the sample.
The current continues to increase until the ferromagnetic region enters the normal state.
The peak in the total current density is therefore not due to a sudden onset of new physics, but rather because the continuous build up of current comes to an end as superconductivity is destroyed.
We also note that the region following the partial collapse of the order parameter and up is complicated by the presence of vortex like currents inside the ferromagnetic bulk region.
See for example Fig.~\ref{Figure:J_77_block_para}.
For strong enough Zeeman term the currents becomes localized on the left and right edges, similarly to what is predicted using a Ginzburg-Landau model in Appendix \ref{Appendix:Ginzburg_Landau}.

\subsection{Wire with magnetic moments along wire}
\begin{figure}
\includegraphics[width=245pt]{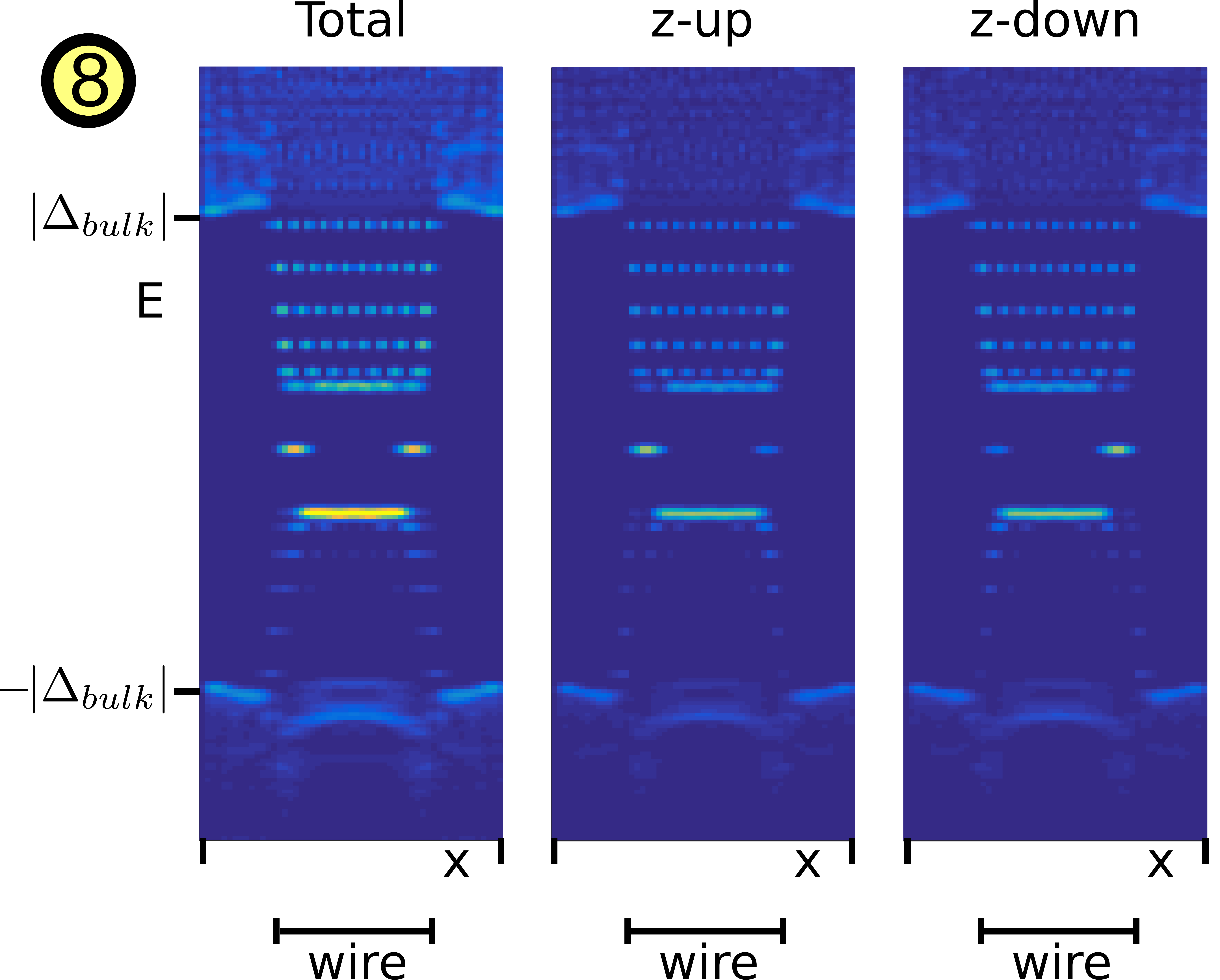}
\caption{(Color online.) Same as in Fig.~\ref{Figure:LDOS}, but for ferromganetic wire magnetized along the wire ($\hat{\mathbf{n}} = \hat{\mathbf{x}}$ and wire along $\hat{\mathbf{x}}$).
Note that the Majorana fermions are spin-polarized along the $z$-axis rather than the $x$-axis.
}
\label{Figure:LDOS_w2_para}
\end{figure}

We now consider a wire with the magnetic moments along the wire.
This is the dimensional reduction of the block in the previous section, where the width of the block has been set to one along the $y$-direction.
This system can, just like the wire in Sec.~\ref{Subsection:Wire_with_Zeeman_term_perpendicular_to_surface}, be described by a Hamiltonian of the form in Eq.~\eqref{Equation:Bulk_Hamiltonian_1D}, if the substitution $\sigma_z \rightarrow \sigma_x$ is made.
This substitution presents no problem because the two Pauli matrices in Eq.~\eqref{Equation:Bulk_Hamiltonian_1D} remain orthogonal, and the topological properties therefore remain the same.
It is in fact appropriate to think of the substitution as a rotation of spin indices around the $y$-axis; $\sigma_x,\sigma_z \rightarrow -\sigma_z, \sigma_x$.
In Fig.~\ref{Figure:LDOS_w2_para} we plot the LDOS and spin-polarized LDOS using the $z$-axis as spin-polarization axis.
It is clear that the Majorana fermions spin-polarization that previously showed up in the $x$-component is now present in the $z$-component.
Moreover, direct comparison to Fig.~\ref{Figure:LDOS} reveals that the $x$-up component goes into the $z$-down components and vice versa, exactly as implied by the rotation of spin-indices.

\begin{figure}
\includegraphics[width=245pt]{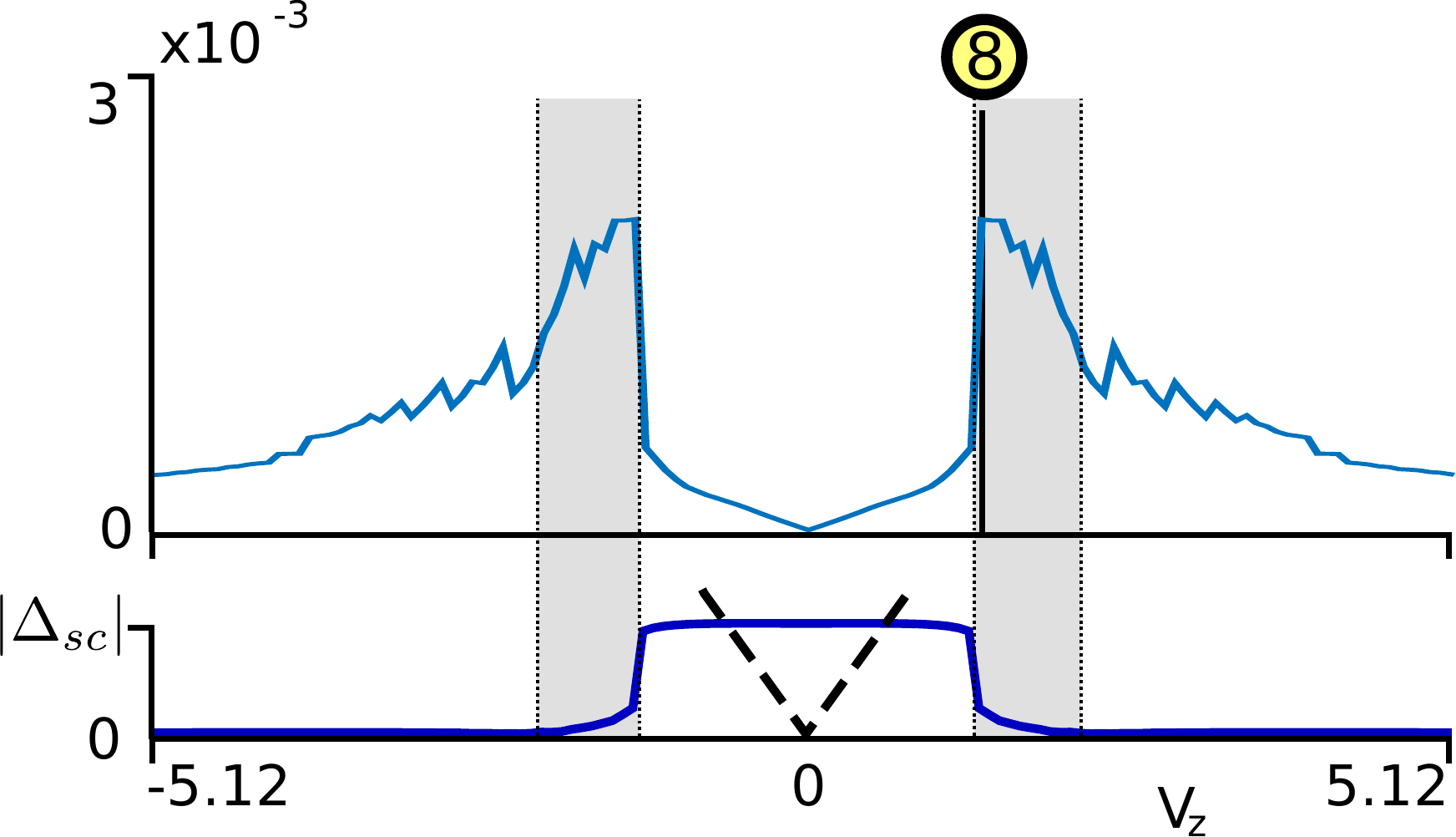}
\caption{(Color online.) Same as in Fig.~\ref{Figure:D_Jmax_block_perp}, but for a ferromagnetic wire magnetized along the wire ($\hat{\mathbf{n}} = \hat{\mathbf{x}}$ and wire along $\hat{\mathbf{x}}$).
The topologically nontrivial region has been identified using the discontinuous onset of strong edge currents, as well as the appearance of Majorana fermions.
}
\label{Figure:D_Jmax_w2_para}
\end{figure}

Finally, the maximum current density and the value of the order parameter at the wires midpoint are shown in Fig.~\ref{Figure:D_Jmax_w2_para}.
The strong discontinuity in the maximum current density is correlated with the appearance of Majorana fermions.
We also mention that the current pattern for the wire closely resembles that for the ferromagnetic block in Fig.~\ref{Figure:J_73_block_para}.
The current flows through the wire perpendicular to the direction of the wire, and the two patterns are so similar that we refrain from showing it in a figure of its own.

\subsection{Wire with in plane magnetic moments perpendicular to wire}
\begin{figure}
\includegraphics[width=245pt]{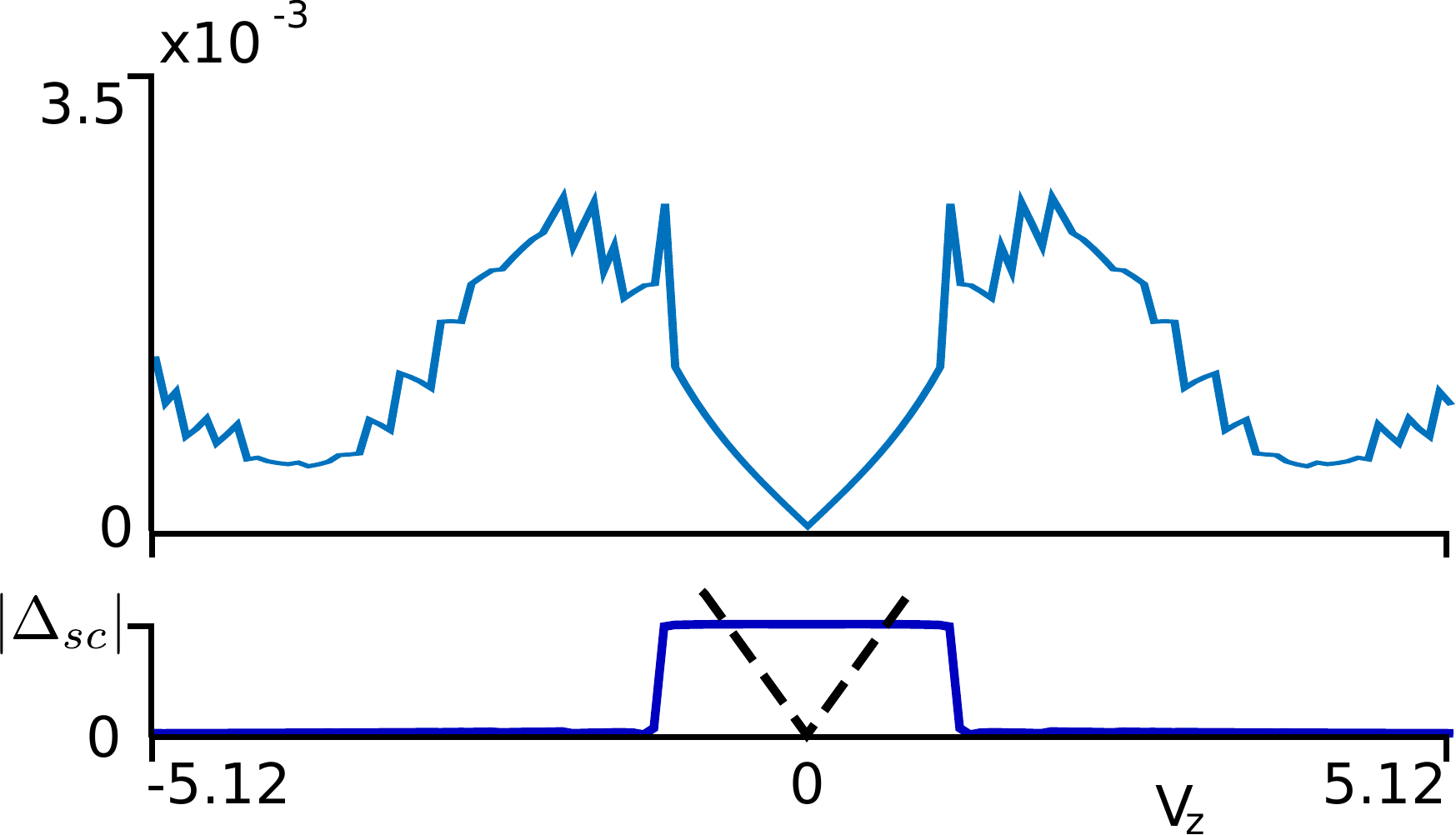}
\caption{(Color online.) Same as in Fig.~\ref{Figure:D_Jmax_block_perp}, but for wire with in-plane magnetization perpendicular to wire ($\hat{\mathbf{n}} = \hat{\mathbf{x}}$ and wire along $\hat{\mathbf{y}}$).
}
\label{Figure:D_Jmax_w1_para}
\end{figure}

The final setup we consider is a wire with in plane magnetic moments perpendicular to the wire.
Similarly as above, this system can be obtained by making the width of the ferromagnetic block in Sec.~\ref{Subsection:Block_with_Zeeman_term_parallel_to_surface} equal to one in the $x$-direction.
However, because the dimension is reduced in a different direction, the substitution of Pauli matrices needed to obtain an effective equation of the form in Eq.~\ref{Equation:Bulk_Hamiltonian_1D} is different.
This time the substitution $k_x, \sigma_y, \sigma_z \rightarrow k_y, -\sigma_x, \sigma_x$ is required, and the two Pauli matrices for the spin-orbit and Zeeman term become the same.
The consequence is that the system has no topologically nontrivial phase, and no Majorana fermions appear.
The maximum current density and order parameter at the midpoint of the wire is presented in Fig.~\ref{Figure:D_Jmax_w1_para}.

\section{Summary}
We have studied ferromagnetic impurities on top of a conventional $s$-wave superconductor with Rashba spin-orbit interaction.
We find a close relation between the topological properties of 2D blocks and 1D wires of the ferromagnetic impurities, as well as the physics of Yu-Shiba-Rusinov states of 0D point impurities.\cite{arXiv:1505.01672}

When the magnetic moments are directed perpendicular to the surface, both the 2D block and 1D wire has topologically nontrivial phases.
For a 2D block the nontrivial phase gives rise to spin-polarized edge currents.
Interestingly, the total persistent current flows in opposite direction to the current contribution from the topologically protected gap-crossing edge states.
This means that different experimental probes will observe different current directions, depending on if they measure the total persistent current or only that carried by low-energy excitations.
The Majorana fermions that appear at the end points of 1D wires are spin-polarized, and their spin-polarization direction are directly related to the spin-polarized edge currents of the 2D block.
The spin-polarization of the Majorana fermions is along the direction of the wire.

When the magnetic moments are parallel to the surface the ferromagnetic block does not have a topologically nontrivial phase.
However, 1D wires can still have a topologically nontrivial phase, depending on if the wire is aligned with the magnetic moments or not.
When the wire is parallel with the magnetic moments, a topologically nontrivial phase exist and the corresponding Majorana fermions are spin-polarized along the $z$-axis.
For both the ferromagnetic block and the two wire directions, a substantial amount of current flows through the bulk, as long as superconductivity is present in the ferromagnetic region.

\section{Acknowledgments}
We thank G.~Volovik, M.~Eschrig, Y.~Kedem, C.~Triola, and T.~Ojanen for useful discussions. This work was supported by the Swedish Research Council (Vetenskapsr\aa det), the G\"oran Gustafsson Foundation, and the Swedish Foundation for Strategic Research (SSF) (K.B. and A.B.-S.) and the European Research Council (ERC) DM-321031 and the US DOE BES E304 (S.S.P. and A.V.B.).

\appendix
\section{Ginzburg Landau approach}
\label{Appendix:Ginzburg_Landau}
\begin{figure*}
\centering	
(a) \includegraphics[height=0.25\linewidth]{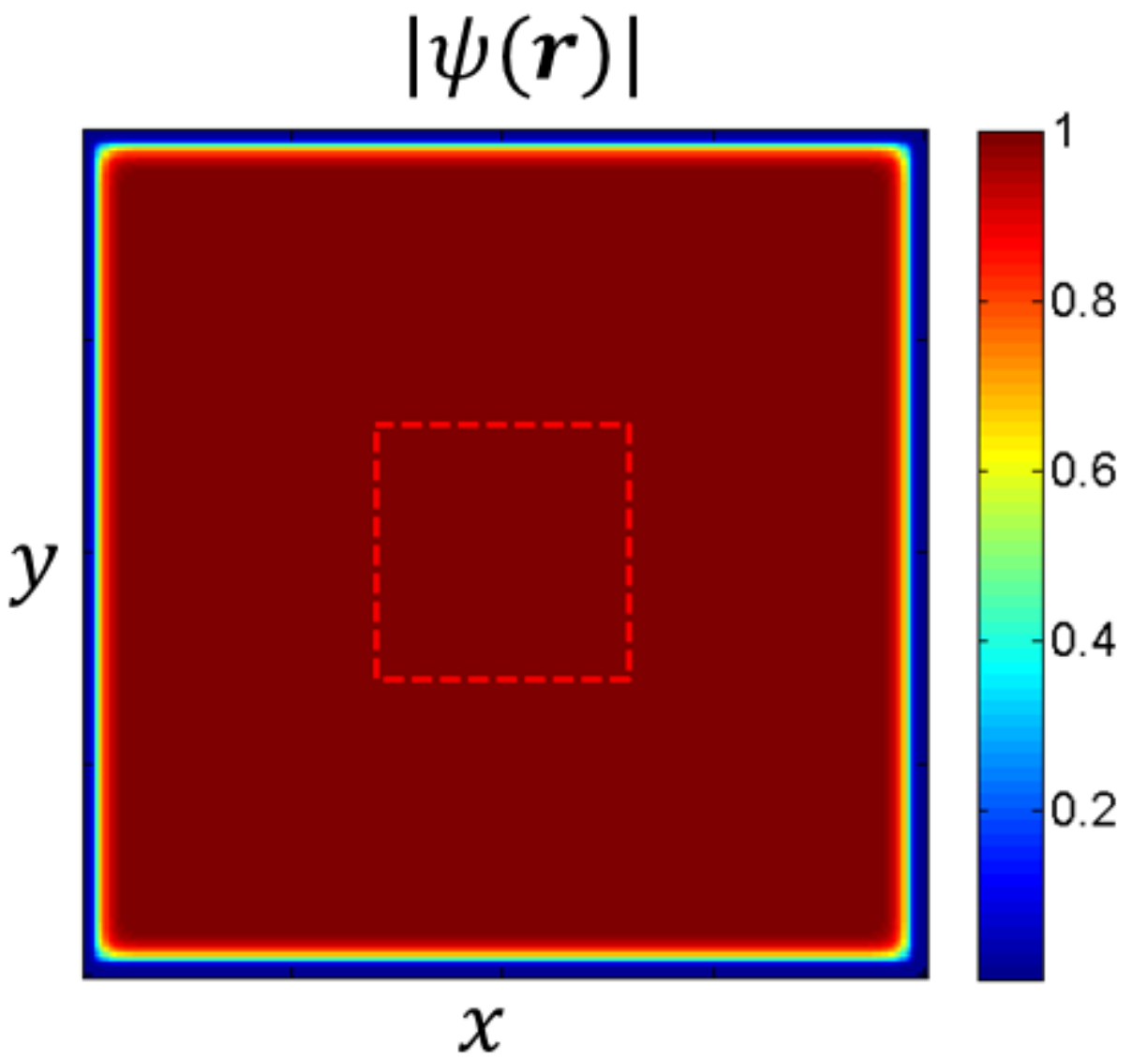} (b) \includegraphics[height=0.25\linewidth]{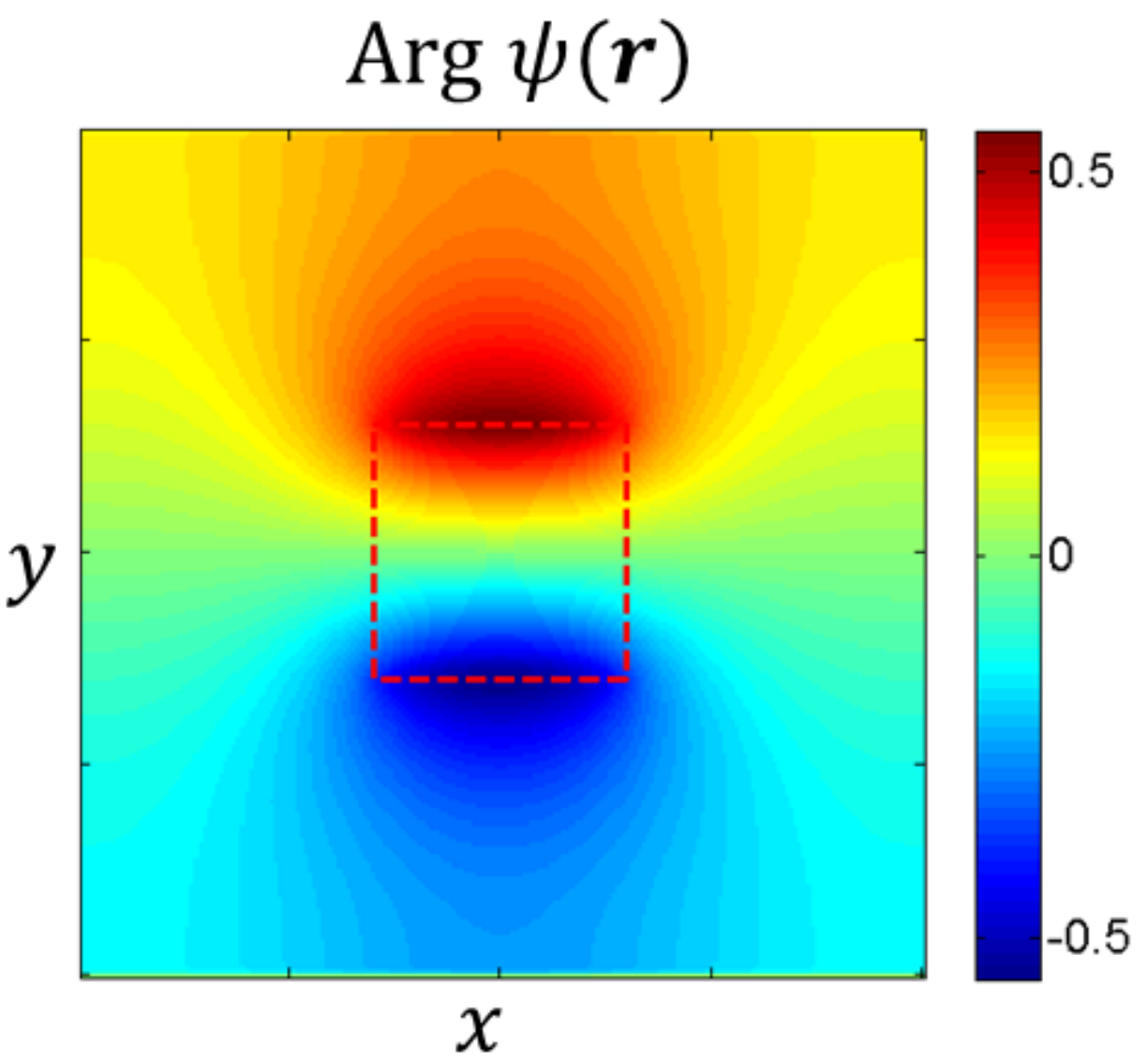} (c) \includegraphics[height=0.25\linewidth]{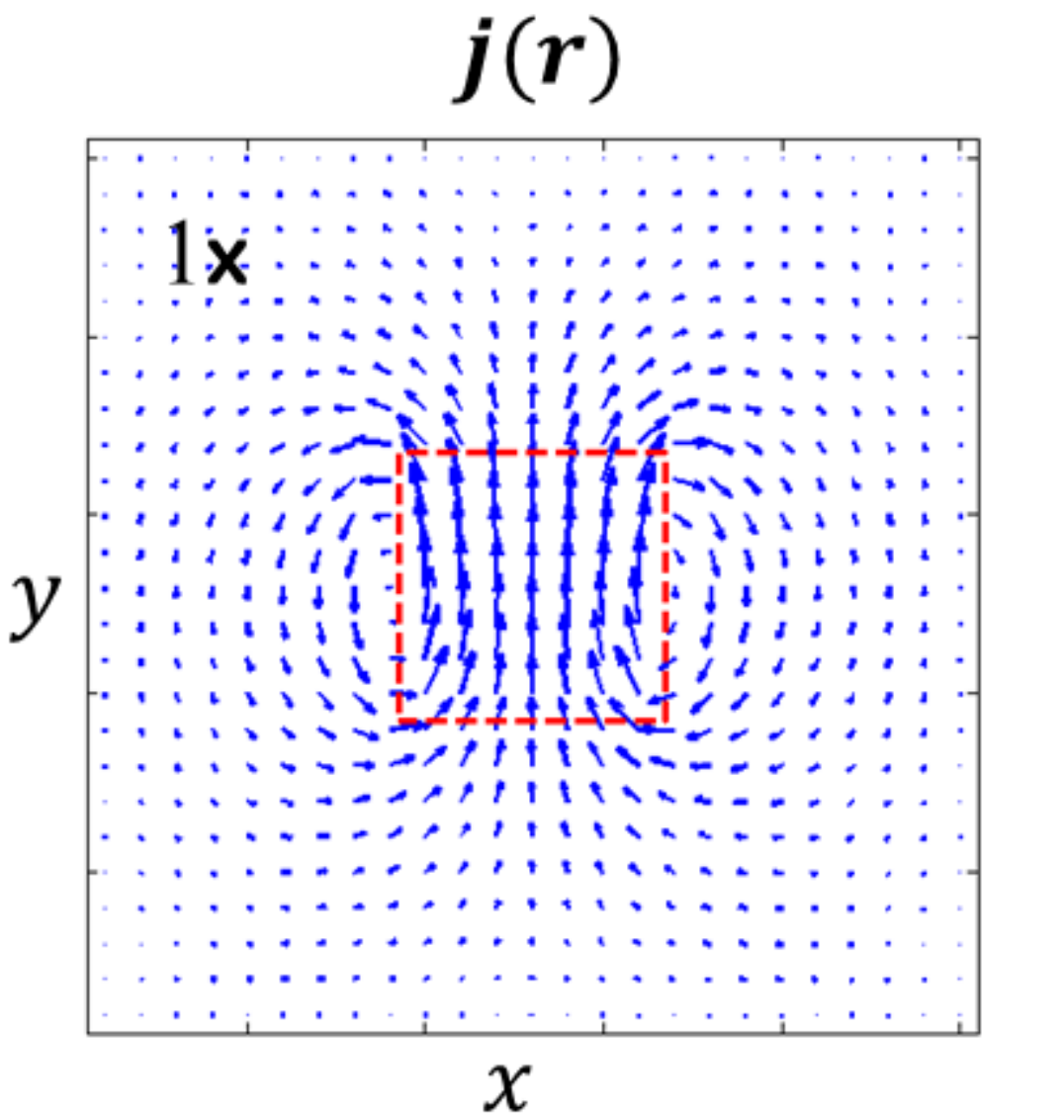} \\  	
(d) \includegraphics[height=0.25\linewidth]{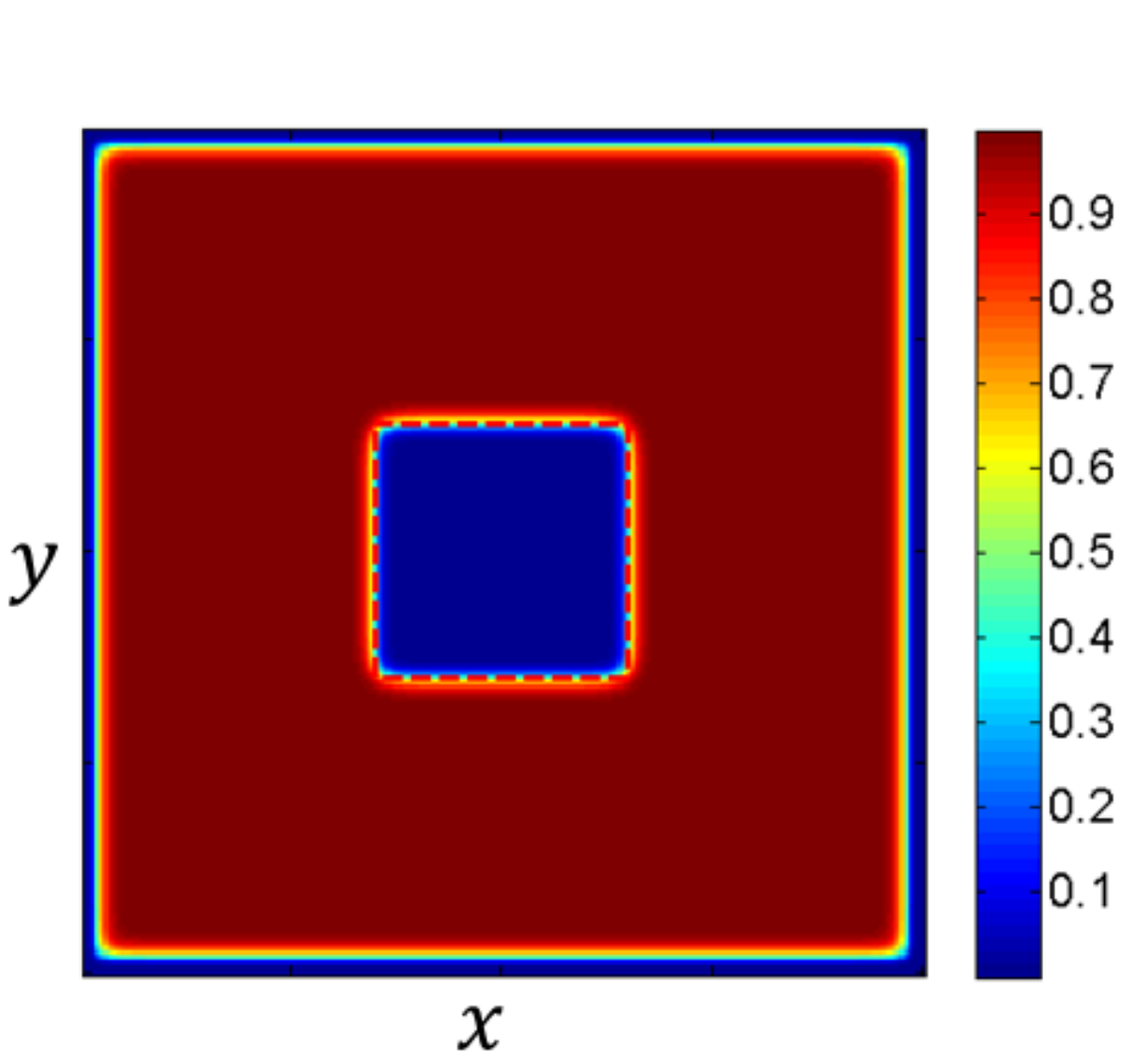} (e) \includegraphics[height=0.25\linewidth]{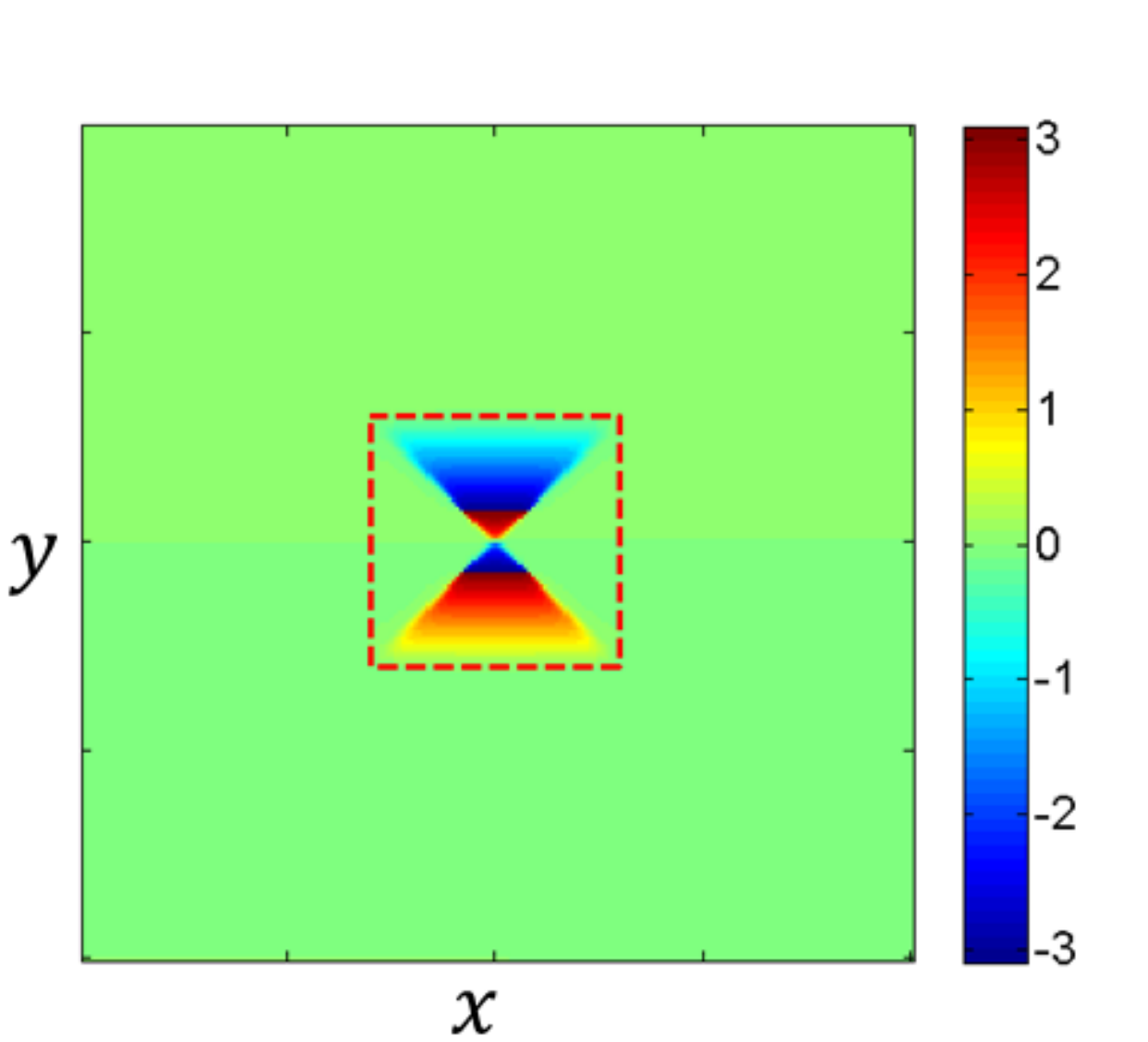} (f) \includegraphics[height=0.25\linewidth]{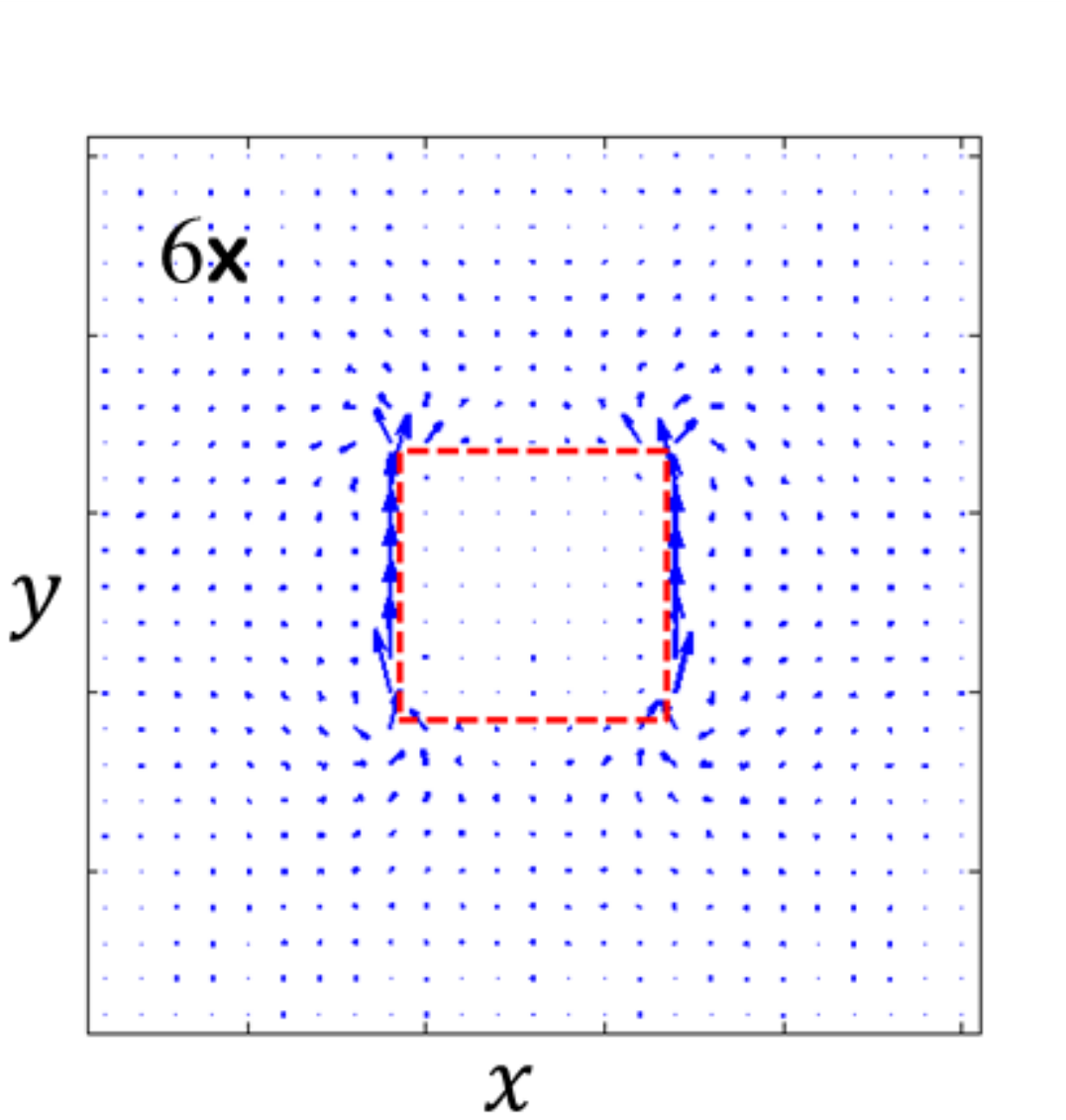} 
\caption{(Color online.)(a, d) Order parameter magnitude $|\psi|$, (b, e) phase Arg$\,\psi$, and currents calculated using Ginzburg-Landau theory. The position of the ferromagnetic block is indicated with a red dashed rectangle. The ferromagnetic vector is in-plane $\bm S = S\hat{\bm x}$. Top row corresponds to $S=0.9$ and bottom row to $S = 2$. The magnitude of the currents in panel (f) is six-fold magnified compared to panel (c) as indicated by the magnification ratio shown in these panels.}
\label{Figure:GL}
\end{figure*}
To complement the self-consistent calculations on a lattice, we here also provide the results of a  Ginzburg-Landau (GL) treatment. The GL method gives an approximate description of the system, but is unable to capture microscopic details such as spin-polarization, edge states, density of states etc.. Nevertheless, it may be interesting to compare the GL results with the exact numerical diagonalization discussed throughout the paper. We consider the following dimensionless GL free energy functional
\begin{align}
	F = &\int\, d^2r\,\,  \left[ \frac{1}{2}\left( |\psi|^2-1 \right)^2 - g_{\psi}\,\frac{S^2(\bm r)}{2} +a(r)|\psi|^2\right.	\label{GL}  \\
	& \left. + b|\bm\nabla\psi|^2+ \,\frac{ic}{2}\left( \psi^* \bm\nabla\psi-\bm\nabla\psi^* \psi \right)\cdot(\hat {\bm z}\times \bm S)\right],  \nonumber
\end{align}
Here, $\psi(\bm r)$ describes the superconducting order parameter in a 2D box, i.e. $\bm r = (x,y)$ and ${-L/2<x,y<L/2}$, where $L=1$. The magnetic block of size $d = 0.3L$ is placed in the center with the magnetization along the $x$ axis, i.e. $S(\bm r) = S\hat{\bm x}$ for $-d/2<x,y<d/2$. The ferromagnetic vector $S$ is analogous to the Zeeman term $V_z$ used throughout the main text. The first term in Eq.~(\ref{GL}) is conventional and favors the superconducting order at $|\psi|=1$, whereas the second term describes the diamagnetic Pauli energy disfavoring superconducting order. The superconducting order is destroyed when the Pauli energy exceeds the the superconducting condensation energy (the so-called Chandrasekhar-Clogston limit), which occurs for $S>S_p=1$ in Eq.~(\ref{GL}). In order to smoothly interpolate  the Pauli term between the superconducting and metallic phases we introduced a sigmoid function
\begin{equation}
      g_\psi = \frac{2}{e^{v|\psi|^2}+1},\quad v = 5.
\end{equation}
Note that $g_\psi\approx 1$ ($0$) in the absence (presence) of the superconductivity, i.e. if $\psi \approx 0$ ($\psi \neq 0$). To avoid possible spurious numerical effects, we also add a boundary term with a large coefficient $a(\bm r)\sim 100$, which destroys superconductivity at the boundary of the 2D box.  The terms on the second line of Eq.~(\ref{GL}) describe the kinetic energy. The term proportional to $b=0.0001$ describes the conventional superfluid kinetic energy, which defines the characteristic coherence length $\xi \sim \sqrt{b} = 0.01$. 
The last term, proportional to $c = 0.003$, describes the magnetoelectric coupling\cite{arXiv:1505.01672} between the current and magnetization $S$. 

We search for the minimum of the free energy~(\ref{GL}) numerically. We employ the gradient descent method in which the order parameter evolves in imaginary time $\tau$ as
\begin{equation}
	\frac{d\psi}{d\tau} = -\frac{\delta F}{\delta \psi}. \label{gradDescent}
\end{equation}
Using Eq.~(\ref{gradDescent}), the order parameter is updated iteratively as $\psi_{\rm new} = \psi_{\rm old}-d\tau\frac{\delta F[\psi_{\rm old}]}{\delta \psi}$. The time evolution stops for the solution $\psi(\bm r)$ satisfying $\frac{\delta F[\psi]}{\delta \psi}=0$, which, thus, minimizes the free energy. 

In Fig.~\ref{Figure:GL} we plot the order parameter magnitude $|\psi|$, phase Arg$\,|\psi|$, and the current
\begin{equation}
	\bm j = ib \left( \psi^* \bm\nabla\psi-\bm\nabla\psi^* \psi \right)+c|\psi|^2\,(\hat {\bm z}\times \bm S).  
	\label{cur}
\end{equation}
The panels in the top (bottom) row correspond to the field $S = 0.9$ ($S = 2$), which is below (above) the Pauli limit $S_p = 1$. For the smaller field (top row), the superconducting order $|\psi|$ is uniform, whereas the superconducting phase is non-uniform and is distributed in such a way that the total current is continuous. In contrast, for the large field (bottom row), the superconducting order is destroyed in the region covered with the ferromagnetic block. The superconducting phase is strongly uniform  outside the ferromagnetic block. The current is suppressed everywhere, except in small vicinity of the ferromagnetic block.

In general, the GL modeling agrees qualitatively with those discussed in Section~\ref{Subsection:Block_with_Zeeman_term_parallel_to_surface}: the current grows linearly with the magnetic field until superconductivity is destroyed. At the same time, we are not able to capture the non-linear peak of the current shown in Fig.~\ref{Figure:D_Jmax_block_para} as well as the appearance of the vortex structure in Fig.~\ref{Figure:J_77_block_para}. 

\section{Derivation current expressions}
\label{Appendix:Derivation_of_expressions_for_currents}
\subsection{Introduction}
Here we derive the general expressions for calculating the currents in an arbitrary spin-basis.
The implementation of the numerical calculations are based on the algorithm described in Sec.~\ref{Section:Summary_of_method}.
The method itself is fairly straightforward.
However, the derivation is complicated by the fact that the spin-polarized currents are not necessarily calculated in the same basis as the Hamiltonian.
A short summary of the appendix follows:\\

\noindent \textit{\ref{Section:Notation}: Notation}\\
\indent Introduction of useful notation and derivations of important identities.\\

\noindent \textit{\ref{Section:Generalized_current_operators}: Generalized current operators}\\
\indent The main problem is formulated.
General sink-source and current operators are defined.
In summary, there are three important terms for any given spin: The sink-source term $\hat{S}_s$, the current $\hat{J}_{\sigma\sigma}^{b}$ on bond $b$, and the spin-flipping current $\hat{J}_{\sigma\bar{\sigma}}^{b}$ on $b$, which together satisfies
\begin{align}
	\frac{d\hat{\rho}_{\sigma}}{dt} = \hat{S}_s - \sum_{b}\left(\hat{J}_{\sigma\sigma} + \hat{J}_{\sigma\bar{\sigma}}\right).
\end{align}
To derive these operators, we need to find a certain set of coefficients $a_{s+b,\eta,s,\eta'}^{\lambda\lambda'\kappa\kappa'}$.\\

\noindent \textit{\ref{Section:Arbitrary basis}: Arbitrary basis}\\
\indent General expressions  are derived for how to calculate the $a_{s+b,\eta,s,\eta'}^{\lambda\lambda'\kappa\kappa'}$ coefficients in an arbitrary basis from the knowledge of those same coefficients in the basis of the Hamiltonian.\\

\noindent \textit{\ref{Section:Calculate_current}: Calculate sink-sources and currents}\\
\indent Expressions for the expectation values of the sink-source and current operators are derived.\\

\noindent \textit{\ref{Section:Summary_of_method}: Summary of method}\\
\indent An algorithmic summary of the method is provided.\\

\noindent \textit{\ref{Section:Tabulating_relevant_as}: Tabulating relevant $a$'s}\\
\indent The exact number of $a_{s+b,\eta,s,\eta'}^{\lambda\lambda'\kappa\kappa'}$ that needs to be evaluated in the basis of the Hamiltonian are determined, and a useful way of tabulating them is introduced.\\

\noindent \textit{\ref{Section:Evaluating_tables}: Evaluating tables}\\
\indent Tables are evaluated for the following types of terms in the Hamiltonian: chemical potential, kinetic, Zeeman, and Rashba spin-orbit interaction.
The total table for a Hamiltonian containing all these terms is also provided.\\

\noindent \textit{\ref{Section:Note_on_superconductivity}: Note on superconductivity}\\
\indent It is shown that the introduction of an $s$-wave superconducting term in the Hamiltonian does not modify the method described in the rest of this appendix.\\

\subsection{Notation}
\label{Section:Notation}
Here we introduce useful notation and show relations which will be important later on.
This section is purely algebraic, and no physical motivation will be provided.
The usefulness of the expressions will only become clear once general currents are treated in Section \ref{Section:Generalized_current_operators}.

\subsubsection{Bond decomposition of commutators}
Consider a general quadratic Hamiltonian
\begin{align}
	\label{Equation:Hamiltonian_decomposition}
	H =& \sum_{\alpha\lambda\lambda'} h_{\lambda\lambda'}^{\alpha},
\end{align}
where
\begin{align}
	h_{\lambda\lambda'}^{\alpha} =& \sum_{ij}f_{i\lambda j\lambda'}^{\alpha}c_{i\lambda}^{\dagger}c_{j\lambda'}.
\end{align}
Here $i,j$ are site indices, $\lambda,\lambda'$ are spin indices, and $\alpha$ is an index distinguishing between different terms in the Hamitlonian such as kinetic, chemical potential, Zeeman, etc..

We now consider expressions of the form
\begin{widetext}
\begin{align}
	\label{Equation:Commutator_decomposed}
	\frac{i}{\hbar}\left[h_{\lambda\lambda'}^{\alpha}, c_{s\kappa}^{\dagger}c_{s\kappa'}\right] =& \frac{i}{\hbar}\sum_{ij}f_{i\lambda j\lambda'}^{\alpha}\left(\delta_{js}\delta_{\lambda'\kappa}c_{i\lambda}^{\dagger}
	c_{s\kappa'} - \delta_{is}\delta_{\lambda\kappa'}c_{s\kappa}^{\dagger}c_{j\lambda'}\right)\nonumber\\
	=& \frac{i}{\hbar}\sum_{b}\left(f_{s+b,\lambda,s\lambda'}^{\alpha}\delta_{\lambda'\kappa}c_{s+b,\lambda}^{\dagger}c_{s\kappa'} - f_{s\lambda,s+b,\lambda'}^{\alpha}\delta_{\lambda\kappa'}c_{s\kappa}^{\dagger}c_{s+b,\lambda'}\right).
\end{align}
\end{widetext}
The summation in the last expression is said to run over all bonds $b$ connected to site $s$, where a bond between site $s$ and $s+b$ is said to exist if the Hamiltonian contains a product of creation and annihilation operators with these two indices.
In particular, the set of bonds has to be understood to include the bond between a site and itself if any on-site terms are included in the Hamiltonian.
Using Eq.~\eqref{Equation:Commutator_decomposed}, it is possible to decompose the commutator for the complete Hamiltonian as
\begin{widetext}
\begin{align}
	\frac{i}{\hbar}\left[H, c_{s\kappa}^{\dagger}c_{s\kappa'}\right] =& \frac{i}{\hbar}\sum_{\alpha\lambda\lambda'}\left[h_{\lambda\lambda'}^{\alpha}, c_{s\kappa}^{\dagger}c_{s\kappa'}\right]\nonumber\\
		=& \frac{i}{\hbar}\sum_{b\lambda}\left(\left(\sum_{\alpha}f_{s+b,\lambda,s\kappa}^{\alpha}\right)
		c_{s+b,\lambda}^{\dagger}c_{s\kappa'} - \left(\sum_{\alpha}f_{s\kappa',s+b,\lambda}^{\alpha}\right)
		c_{s\kappa}^{\dagger}c_{s+b,\lambda}\right).
\end{align}
\end{widetext}
Defining
\begin{align}
	\label{Equation:Definition_a}
	a_{i,\eta,j,\eta'}^{\lambda\lambda'\kappa\kappa'} =& \delta_{\lambda\eta}\delta_{\kappa'\eta'}\delta_{\lambda'\kappa}\frac{i}{\hbar}\sum_{\alpha}f_{i,\lambda,j,\lambda'}^{\alpha},\nonumber\\
	b_{i,\eta,j,\eta'}^{\lambda\lambda'\kappa\kappa'} =& \delta_{\kappa\eta}\delta_{\lambda'\eta'}\delta_{\lambda\kappa'}\frac{i}{\hbar}\sum_{\alpha}f_{i,\lambda,j,\lambda'}^{\alpha},
\end{align}
the expression can further be written as
\begin{widetext}
\begin{align}
	\label{Equation:Bond_decomposition}
		\frac{i}{\hbar}\left[H, c_{s\kappa}^{\dagger}c_{s\kappa'}\right]
		=& \sum_{b\lambda\lambda'\eta\eta'}\left(a_{s+b,\eta,s,\eta'}^{\lambda\lambda'\kappa\kappa'}c_{s+b,\eta}^{\dagger}c_{s\eta'} - b_{s,\eta,s+b,\eta'}^{\lambda\lambda'\kappa\kappa'}c_{s\eta}^{\dagger}c_{s+b,\eta'}\right)\nonumber\\
		=& \sum_{b\lambda}\left(a_{s+b,\lambda,s,\kappa'}^{\lambda\kappa\kappa\kappa'}c_{s+b,\lambda}^{\dagger}c_{s\kappa'} - b_{s,\kappa,s+b,\lambda}^{\kappa'\lambda\kappa\kappa'}c_{s\kappa}^{\dagger}c_{s+b,\lambda}\right).
\end{align}
\end{widetext}

\subsubsection{Useful expressions}
The Hamiltonian is Hermitian if $H = H^{\dagger}$, which using Eq.~\eqref{Equation:Hamiltonian_decomposition} implies
\begin{align}
	\sum_{\alpha\lambda\lambda' ij}f_{i\lambda j\lambda'}^{\alpha}c_{i\lambda}^{\dagger}c_{j\lambda'} = \sum_{\alpha\lambda\lambda' ij}f_{i\lambda j\lambda'}^{\alpha *}c_{j\lambda'}^{\dagger}c_{i\lambda}.
\end{align}
Identification of coefficients leads to the conclusion that $\sum_{\alpha}f_{i\lambda j\lambda'}^{\alpha*} = \sum_{\alpha}f_{j\lambda' i\lambda}^{\alpha}$, and using Eq.~\eqref{Equation:Definition_a} we have
\begin{align}
	\label{Equation:Relations_ab}
	b_{i\eta j\eta'}^{\lambda\lambda'\kappa\kappa'} =& -a_{j\eta'i\eta}^{\lambda'\lambda\kappa'\kappa*}.
\end{align}
It is therefore possible to rewrite Eq.~\eqref{Equation:Bond_decomposition} as
\begin{widetext}
\begin{align}
	\label{Equation:Bond_decomposition_reduced}
	\frac{i}{\hbar}\left[H, c_{s\kappa}^{\dagger}c_{s\kappa'}\right] =& \sum_{b\lambda\lambda'\eta\eta'}\left(a_{s+b,\eta,s,\eta'}^{\lambda\lambda'\kappa\kappa'}c_{s+b,\eta}^{\dagger}c_{s\eta'} + a_{s+b,\eta',s,\eta}^{\lambda'\lambda\kappa'\kappa*}c_{s\eta}^{\dagger}c_{s+b,\eta'}\right)\nonumber\\
		=& \sum_{b\lambda}\left(a_{s+b,\lambda,s,\kappa'}^{\lambda\kappa\kappa\kappa'}c_{s+b,\lambda}^{\dagger}c_{s\kappa'} + a_{s+b,\lambda,s,\kappa}^{\lambda\kappa'\kappa'\kappa*}c_{s\kappa}^{\dagger}c_{s+b,\lambda}\right),
\end{align}
\end{widetext}
which eliminates the need to calculate $b$ coefficients independently.
We also note from Eq.~\eqref{Equation:Definition_a} that $a_{i\eta i\eta}^{\lambda\lambda\kappa\kappa} = b_{i\eta i\eta}^{\lambda\lambda\kappa\kappa}$, which together with Eq.~\eqref{Equation:Relations_ab} gives
\begin{align}
	\label{Equation:Diagonal_a_is_zero}
	a_{i\eta i\eta}^{\lambda\lambda\kappa\kappa} = 0.
\end{align}
Using $\bar{\kappa}'$ and $\bar{\eta}'$ to denote the spins which are opposite to $\kappa'$ and $\eta'$, respectively, we also find from Eq.~\eqref{Equation:Definition_a} that
\begin{align}
	\label{Equation:Relations_aa}
	a_{i,\eta,j,\eta'}^{\lambda\lambda'\kappa\kappa'} = a_{i,\eta,j,\bar{\eta}'}^{\lambda\lambda'\kappa\bar{\kappa}'}
\end{align}

\subsubsection{Obtaing $a$'s through projection}
In Eq.~\eqref{Equation:Definition_a} explicit expressions for how to calculate $a_{i\eta j\eta'}^{\lambda\lambda'\kappa\kappa'}$ were provided.
It will also be useful to have a way of extracting coefficients from a general quadratic expression, also when the explicit expression for these coefficients are not known.
To this end we consider a general quadratic expression
\begin{align}
	e = \sum_{\mu\nu}a_{\mu\nu}c_{\mu}^{\dagger}c_{\nu},
\end{align}
where $\mu$ and $\nu$ are some arbitrary set of indices.
It is clear that
\begin{align}
	\langle 0|c_{\rho}ec_{\rho'}^{\dagger}|0\rangle = a_{\rho\rho'}.
\end{align}
Applying this to Eq.~\eqref{Equation:Bond_decomposition_reduced} we can for $b\neq0$ write
\begin{align}
	\label{Equation:Projector_on_bond}
	\langle 0|c_{s+b,\eta}\frac{i}{\hbar}\left[H, c_{s\kappa}^{\dagger}c_{s\kappa'}\right]c_{s\eta'}^{\dagger}|0\rangle =& \sum_{\lambda\lambda'}a_{s+b,\eta,s,\eta'}^{\lambda\lambda'\kappa\kappa'} = a_{s+b,\eta,s,\eta'}^{\eta\kappa\kappa\kappa'},
\end{align}
while for $b = 0$ we have
\begin{align}
	\langle 0|c_{s\eta}\frac{i}{\hbar}\left[H, c_{s\kappa}^{\dagger}c_{s\kappa'}\right]c_{s\eta'}^{\dagger}|0\rangle =& a_{s,\eta,s,\eta'}^{\eta\kappa\kappa\kappa'} + a_{s,\eta',s,\eta}^{\eta'\kappa'\kappa'\kappa*},
\end{align}
which can also be written as
\begin{widetext}
\begin{align}
	\label{Equation:Projector_on_site}
	\langle 0|c_{s\eta}\frac{i}{\hbar}\left[H, c_{s\kappa}^{\dagger}c_{s\kappa'}\right]c_{s\eta'}^{\dagger}|0\rangle =&
	\left\{\begin{array}{ccc}
		0												&	\textrm{if}	& \eta \neq \kappa\textrm{\;and\;} \eta' \neq \kappa',\\
		a_{s\eta s\eta'}^{\eta\kappa\kappa\kappa'}		&	\textrm{if}	& \eta \neq \kappa \textrm{\;and\;} \eta' = \kappa',\\
		a_{s\eta's\eta}^{\eta'\kappa'\kappa'\kappa*}	&	\textrm{if}	& \eta = \kappa \textrm{\;and\;}\eta' \neq \kappa',\\
		a_{s,\eta,s,\eta'}^{\eta\kappa\kappa\kappa'} + a_{s,\eta',s,\eta}^{\eta'\kappa'\kappa'\kappa*}						&	\textrm{if}	& \eta = \kappa \textrm{\;and\;} \eta' = \kappa'.
	\end{array}\right.
\end{align}
\end{widetext}

\subsection{Generalized current operators}
\label{Section:Generalized_current_operators}
With the notation introduced in the previous section, we are now ready to formulate the problem in a way that eventually will allow us to calculate spin-polarized currents in any given basis.
To this end we consider the local density operator for particles of spin $\sigma$ on site $s$, which can be written as $\hat{\rho}_{s\sigma} = c_{s\sigma}^{\dagger}c_{s\sigma}$.
The time rate of change of this operator is
\begin{align}
	\frac{d\hat{\rho}_{s\sigma}}{dt} = \frac{i}{\hbar}\left[H, \hat{\rho}_{s\sigma}\right].
\end{align}
The right hand side of this expression is of the form considered in the previous section, and we therefore know from Eq.~\eqref{Equation:Bond_decomposition_reduced} that it can be written as
\begin{align}
			\frac{d\hat{\rho}_{s\sigma}}{dt} =& \sum_{b\lambda}\left(a_{s+b,\lambda,s,\sigma}^{\lambda\sigma\sigma\sigma}c_{s+b,\lambda}^{\dagger}c_{s\sigma} + a_{s+b,\lambda,s,\sigma}^{\lambda\sigma\sigma\sigma*}c_{s\sigma}^{\dagger}c_{s+b,\lambda}\right)\nonumber\\
		=& \left(a_{s,\sigma,s,\sigma}^{\sigma\sigma\sigma\sigma} + a_{s,\sigma,s,\sigma}^{\sigma\sigma\sigma\sigma*}\right)c_{s\sigma}^{\dagger}c_{s\sigma}\nonumber\\
		&+ a_{s,\bar{\sigma},s,\sigma}^{\bar{\sigma}\sigma\sigma\sigma}c_{s\bar{\sigma}}^{\dagger}c_{s\sigma} + a_{s,\bar{\sigma},s,\sigma}^{\bar{\sigma}\sigma\sigma\sigma*}c_{s\sigma}^{\dagger}c_{s\bar{\sigma}}\nonumber\\
		&+ \sum_{b\neq 0}\left(a_{s+b,\sigma,s,\sigma}^{\sigma\sigma\sigma\sigma}c_{s+b,\sigma}^{\dagger}c_{s\sigma} + a_{s+b,\sigma,s,\sigma}^{\sigma\sigma\sigma\sigma*}c_{s\sigma}^{\dagger}c_{s+b,\sigma}\right)\nonumber\\
		&+ \sum_{b\neq 0}\left(a_{s+b,\bar{\sigma},s,\sigma}^{\bar{\sigma}\sigma\sigma\sigma}c_{s+b,\bar{\sigma}}^{\dagger}c_{s\sigma} + a_{s+b,\bar{\sigma},s,\sigma}^{\bar{\sigma}\sigma\sigma\sigma*}c_{s\sigma}^{\dagger}c_{s+b,\bar{\sigma}}\right),
\end{align}
where $\bar{\sigma}$ is the opposite spin of $\sigma$.
In the last step the on-site terms have been singled out of the expression, and the remaining sum over the bonds have been divided into one part free of spin-flips, and one involving spin-flips.
Using Eq.~\eqref{Equation:Diagonal_a_is_zero} we can further conclude that the first term is zero.
The next two terms converts $\sigma$- and $\bar{\sigma}$-spins into each other on-site.
When considering the density of a single spin species such as $\sigma$, these terms therefore acts as sink and source terms, respectively.
We therefore define the sink-source operator on site $s$ as
\begin{align}
	\label{Equation:Sink_source}
	\hat{S}_{s} =& a_{s,\bar{\sigma},s,\sigma}^{\bar{\sigma}\sigma\sigma\sigma}c_{s\bar{\sigma}}^{\dagger}c_{s\sigma} + a_{s,\bar{\sigma},s,\sigma}^{\bar{\sigma}\sigma\sigma\sigma*}c_{s\sigma}^{\dagger}c_{s\bar{\sigma}}.
\end{align}
The third line describes processes where a $\sigma$ spin jumps between site $s$ and site $s+b$.
It is therefore the current into site $s$ along bond $b$.
We accordingly define the current operator along bond $b$ as
\begin{align}
	\label{Equation:Current}
	\hat{J}_{\sigma\sigma}^{b} =& -\left(a_{s+b,\sigma,s,\sigma}^{\sigma\sigma\sigma\sigma}c_{s+b,\sigma}^{\dagger}c_{s\sigma} + a_{s+b,\sigma,s,\sigma}^{\sigma\sigma\sigma\sigma*}c_{s\sigma}^{\dagger}c_{s+b,\sigma}\right),
\end{align}
where it is implicitly understood that the operator is only defined for actual bonds between different sites.
We have here included an additional minus sign in order to define the direction of the current to be along the bond away from site $s$.
The third term is similarly related to the in and out flow of $\sigma$-spins on site $s$.
However, in contrast to the ordinary currents the spins flip during the transition.
We therefore define the spin-flipping current along bond $b$ as
\begin{align}
	\label{Equation:Current_spin_flip}
	\hat{J}_{\sigma\bar{\sigma}}^{b} =& -\left(a_{s+b,\bar{\sigma},s,\sigma}^{\bar{\sigma}\sigma\sigma\sigma}c_{s+b,\bar{\sigma}}^{\dagger}c_{s\sigma} + a_{s+b,\bar{\sigma},s,\sigma}^{\bar{\sigma}\sigma\sigma\sigma*}c_{s\sigma}^{\dagger}c_{s+b,\bar{\sigma}}\right).
\end{align}

With these definitions we can now write
\begin{align}
	\label{Equation:Density_source_current}
	\frac{d\hat{\rho}_{s\sigma}}{dt} = \hat{S}_{s} - \sum_{b\neq 0}\left(\hat{J}_{\sigma\sigma}^{b} + \hat{J}_{\sigma\bar{\sigma}}^{b}\right).
\end{align}
It is clear that in order to have a complete understanding of the currents in a given basis $\sigma$, it is sufficient to evaluate the three coefficients
\begin{align}
	\label{Equation:Relevant_coefficients}
	\left\{a_{s,\bar{\sigma},s,\sigma}^{\bar{\sigma}\sigma\sigma\sigma},
	a_{s+b,\sigma,s,\sigma}^{\sigma\sigma\sigma\sigma},
	a_{s+b,\bar{\sigma},s,\sigma}^{\bar{\sigma}\sigma\sigma\sigma}\right\}.
\end{align}
These can all be written as
\begin{align}
	a_{s+b,\sigma',s,\sigma}^{\sigma'\sigma\sigma\sigma}.
\end{align}

With knowledge of the eigenstates of the Hamiltonian, the expressions just derived allow us to calculate on-site sink-sources, as well as currents on bonds, using $S_{s} = \langle\hat{S}_s\rangle$ and $J_{\sigma\sigma'}^{b} = \langle\hat{J}_{\sigma\sigma'}^{b}\rangle$, respectively.
However, it is also useful to know the current on a specific site, rather than on the bonds.
If $\mathbf{b}$ is the vector along bond $b$, from $s$ to $s+b$, we also define the two types of currents at site $s$ as
\begin{align}
	\mathbf{J}_{\sigma\sigma'}^{s} = \frac{1}{2}\sum_{b\neq 0}\mathbf{b}J_{\sigma\sigma'}^{b}.
\end{align}
This is a directed average of all currents flowing out from site $s$ and forms a vector field, in contrast to the scalar currents defined on the bonds.

\subsection{Arbitrary basis}
\label{Section:Arbitrary basis}
Although we in principle can proceed to calculate the relevant coefficients in Eq.~\eqref{Equation:Relevant_coefficients} using Eq.~\eqref{Equation:Definition_a}, this requires the Hamiltonian to be expressed in the same basis as $\sigma$.
It would be a nuisance if each time we want to calculate the current in a new basis, the Hamiltonian would need to be rewritten in that basis, commutators be reevaluated, and finally  also the expectation values $\langle\hat{S}_s\rangle$ and $\langle\hat{J}_{\sigma\sigma'}^{b}\rangle$.
It is much better if the current in one basis can be evaluated from the commutators and expectation values calculated in the basis of the original Hamiltonian.

Consider therefore operators with arbitrary spin direction
\begin{align}
	c_{s\sigma} =& b_{\uparrow}c_{s\uparrow} + b_{\downarrow}c_{s\downarrow},\nonumber\\
	c_{s\bar{\sigma}} =& b_{\downarrow}^{*}c_{s\uparrow} - b_{\uparrow}^{*}c_{s\downarrow} \equiv \bar{b}_{\uparrow}c_{s\uparrow} + \bar{b}_{\downarrow}c_{s\downarrow},
\end{align}
where $\sigma$ is some arbitrary spin, and $\bar{\sigma}$ is the opposite spin.
Further, $\uparrow$ and $\downarrow$ here refers to up and down spins in the basis of the Hamiltonian.
If $\sigma$ is along the direction ($\theta,\varphi$), in ordinary polar coordinates relative to the basis of the Hamiltonian, the coefficients can be chosen to be
\begin{align}
	b_{\uparrow}		=& -\bar{b}_{\downarrow}^{*}	= \cos(\frac{\theta}{2}),\nonumber\\
	b_{\downarrow}	=& \bar{b}_{\uparrow}^{*}	=\cos(\frac{\theta}{2})e^{-i\varphi}.
\end{align}
and the relevant commutator can be written as
\begin{align}
	\frac{i}{\hbar}\left[H, c_{s\sigma}^{\dagger}c_{s\sigma}\right] = \sum_{\kappa\kappa'}b_{\kappa}^{*}b_{\kappa'}\frac{i}{\hbar}\left[H, c_{s\kappa}^{\dagger}c_{s\kappa'}\right].
\end{align}
It is clear that the left hand side can be calculated for any spin direction $\sigma$ if the right hand side has been evaluated in the basis of the Hamiltonian.
However, we remember that it is the expansion of the left hand side into $a$ coefficients that allows us to calculate currents.
Assume therefore that the $a$ coefficients in the expansion of the commutators on the right hand side have been evaluated.
We denote these coefficients by a capital $A$, to distinguish them from the $a$ coefficients in the expansion in the basis of the expression on the left hand side.

We will now show how $a$ coefficients in any basis can be calculated from a knowledge of the $A$ coefficients.
To this end, we remind ourselves that we in fact are only interested in calculating the three $a$'s in Eq.~\eqref{Equation:Relevant_coefficients}.
Further, using Eq.~\eqref{Equation:Projector_on_bond} and \eqref{Equation:Projector_on_site}, these can be written as
\begin{widetext}
\begin{align}
	a_{s,\bar{\sigma},s,\sigma}^{\bar{\sigma}\sigma\sigma\sigma} =& \langle 0|c_{s,\bar{\sigma}}\frac{d\hat{\rho}_{\sigma}}{dt}c_{s\sigma}^{\dagger}|0\rangle\nonumber\\
		=& \sum_{\kappa\kappa'\kappa''\kappa'''}
			b_{\kappa}^{*}b_{\kappa'}\bar{b}_{\kappa''}b_{\kappa'''}^ {*}\langle 0|c_{s,\kappa''}\frac{i}{\hbar}\left[H, c_{s\kappa}^{\dagger}c_{s\kappa'}\right]c_{s\kappa'''}^{\dagger}|0\rangle,\nonumber\\
	a_{s+b,\sigma,s,\sigma}^{\sigma\sigma\sigma\sigma} =& \langle 0|c_{s+b,\sigma}\frac{d\hat{\rho}_{\sigma}}{dt}c_{s\sigma}^{\dagger}|0\rangle\nonumber\\
		=& \sum_{\kappa\kappa'\kappa''\kappa'''}b_{\kappa}^{*}
			b_{\kappa'}b_{\kappa''}b_{\kappa'''}^{*}\langle 0|c_{s+b,\kappa''}\frac{i}{\hbar}\left[H, c_{s\kappa}^{\dagger}c_{s\kappa'}\right]c_{s\kappa'''}^{\dagger}|0\rangle\nonumber\\
	a_{s+b,\bar{\sigma},s,\sigma}^{\bar{\sigma}\sigma\sigma\sigma} =& \langle 0|c_{s+b,\bar{\sigma}}\frac{d\hat{\rho}_{\sigma}}{dt}c_{s\sigma}^{\dagger}|0\rangle\nonumber\\
		=& \sum_{\kappa\kappa'\kappa''\kappa'''}
			b_{\kappa}^{*}b_{\kappa'}\bar{b}_{\kappa''}b_{\kappa'''}^ {*}\langle 0|c_{s+b,\kappa''}\frac{i}{\hbar}\left[H, c_{s\kappa}^{\dagger}c_{s\kappa'}\right]c_{s\kappa'''}^{\dagger}|0\rangle.
\end{align}
\end{widetext}

The two last expressions can be simplified straightforwardly by using Eq.~\eqref{Equation:Projector_on_bond} and then summing over indices appearing in $\delta$-functions in Eq.~\eqref{Equation:Definition_a}, which gives
\begin{align}
	\label{Equation:a_coefficients_relation_1}
	a_{s+b,\sigma,s,\sigma}^{\sigma\sigma\sigma\sigma}
		=& \sum_{\kappa\kappa'\kappa''}
		b_{\kappa}^{*}|b_{\kappa'}|^2 b_{\kappa''}A_{s+b,\kappa'',s,\kappa'}^{\kappa''\kappa\kappa\kappa'},\nonumber\\
	a_{s+b,\bar{\sigma},s,\sigma}^{\bar{\sigma}\sigma\sigma\sigma}
		=& \sum_{\kappa\kappa'\kappa''}
		b_{\kappa}^{*}|b_{\kappa'}|^2 \bar{b}_{\kappa''}A_{s+b,\kappa'',s,\kappa'}^{\kappa''\kappa\kappa\kappa'}.
\end{align}
The on-site term is slightly more complicated due to the special cases in Eq.~\eqref{Equation:Projector_on_site}:
\begin{align}
	\label{Equation:a_coefficients_relation_2}
	a_{s,\bar{\sigma},s,\sigma}^{\bar{\sigma}\sigma\sigma\sigma}	=& \sum_{\kappa\kappa'\kappa''\kappa'''}
		b_{\kappa}^{*}b_{\kappa'}\bar{b}_{\kappa''}b_{\kappa'''}^{*}\left(A_{s,\kappa'',s,\kappa'''}^{\kappa''\kappa\kappa\kappa'} + A_{s,\kappa''',s,\kappa''}^{\kappa'''\kappa'\kappa'\kappa*}\right).
\end{align}

\subsection{Sink-sources and currents}
\label{Section:Calculate_current}
Using Eqs.~\eqref{Equation:Sink_source}-\eqref{Equation:Current_spin_flip} together with Eq.~\eqref{Equation:a_coefficients_relation_1} and Eq.~\eqref{Equation:a_coefficients_relation_2}, we can now write the expectation values for the sink-source and current operators as
\begin{widetext}
\begin{align}
	\langle \hat{S}_{s}\rangle =& \sum_{\kappa\kappa'}\left(a_{s,\bar{\sigma},s,\sigma}^{\bar{\sigma}\sigma\sigma\sigma}
	\bar{b}_{\kappa}^{*}b_{\kappa'}\langle c_{s\kappa}^{\dagger}c_{s\kappa'}\rangle_{H} + a_{s,\bar{\sigma},s,\sigma}^{\bar{\sigma}\sigma\sigma\sigma*}b_{\kappa}^{*}\bar{b}_{\kappa'}\langle c_{s\kappa}^{\dagger}c_{s\kappa'}\rangle_{H}\right),\nonumber\\
	\langle\hat{J}_{\sigma\sigma}^{b}\rangle =& -\sum_{\kappa\kappa'}b_{\kappa}^{*}b_{\kappa'}\left(a_{s+b,\sigma,s,\sigma}^{\sigma\sigma\sigma\sigma}\langle c_{s+b,\kappa}^{\dagger}c_{s\kappa'}\rangle_{H} + a_{s+b,\sigma,s,\sigma}^{\sigma\sigma\sigma\sigma*}\langle c_{s\kappa}^{\dagger}c_{s+b,\kappa'}\rangle_{H}\right),\nonumber\\
	\langle\hat{J}_{\sigma\bar{\sigma}}^{b}\rangle =& -\sum_{\kappa\kappa'}\left(a_{s+b,\bar{\sigma},s,\sigma}^{\bar{\sigma}\sigma\sigma\sigma}\bar{b}_{\kappa}^{*}b_{\kappa'}\langle c_{s+b,\kappa}^{\dagger}c_{s\kappa'}\rangle_{H} + a_{s+b,\bar{\sigma},s,\sigma}^{\bar{\sigma}\sigma\sigma\sigma*}b_{\kappa}^{*}\bar{b}_{\kappa'}\langle c_{s\kappa}^{\dagger}c_{s+b,\kappa'}\rangle_{H}\right).
\end{align}
\end{widetext}
Here also the expectation values on the right hand side have been expanded in the basis of the Hamiltonian, which is indicated by the subscript $H$.
Using that $\langle c_{s\kappa}^{\dagger}c_{s+b,\kappa'}\rangle = \langle c_{s+b,\kappa'}^{\dagger}c_{s\kappa}\rangle^{*}$, this can finally be written as
\begin{align}
	\langle \hat{S}_{s}\rangle =& 2\textrm{Re}\left(a_{s,\bar{\sigma},s,\sigma}^{\bar{\sigma}\sigma\sigma\sigma}
	\sum_{\kappa\kappa'}	\bar{b}_{\kappa}^{*}b_{\kappa'}\langle c_{s\kappa}^{\dagger}c_{s\kappa'}\rangle_{H}\right),\nonumber\\
	\langle\hat{J}_{\sigma\sigma}^{b}\rangle =& -2\textrm{Re}\left(a_{s+b,\sigma,s,\sigma}^{\sigma\sigma\sigma\sigma} \sum_{\kappa\kappa'}b_{\kappa}^{*}b_{\kappa'}\langle c_{s+b,\kappa}^{\dagger}c_{s\kappa'}\rangle_{H}\right),\nonumber\\
	\langle\hat{J}_{\sigma\bar{\sigma}}^{b}\rangle =& -2\textrm{Re}\left(a_{s+b,\bar{\sigma},s,\sigma}^{\bar{\sigma}\sigma\sigma\sigma}\sum_{\kappa\kappa'}\bar{b}_{\kappa}^{*}b_{\kappa'}\langle c_{s+b,\kappa}^{\dagger}c_{s\kappa'}\rangle_{H}\right).
\end{align}

\subsection{Summary of method}
\label{Section:Summary_of_method}
Here we give a short algorithmic summary of the method.
We note that the two first steps only needs to be performed once in the basis of the Hamiltonian.
Repeated application of the subsequent steps then allows for the sink-sources and currents in arbitrary bases to be calculated.

\subsubsection{Calculate expansion coefficients of commutators in the basis of the Hamiltonian}
\begin{align}
	A_{s+b,\eta,s,\eta'}^{\lambda\lambda'\kappa\kappa'} =& \delta_{\lambda\eta}\delta_{\lambda'\kappa}\delta_{\kappa'\eta'}\frac{i}{\hbar}\sum_{\alpha}f_{s+b,\lambda,s,\lambda'}^{\alpha}.
\end{align}
See Sec.~\ref{Section:Evaluating_tables} for already evaluated terms for chemical potential, kinetic, Zeeman, and Rashba spin-orbit interaction.

\subsubsection{Calculate expectation values in the basis of the Hamiltonian}
\begin{align}
	\langle c_{s+b,\kappa}^{\dagger}c_{s\kappa'}\rangle_{H}.
\end{align}

\subsubsection{Choose spin direction}
Choose appropriate coefficients $b_{\uparrow}$ and $b_{\downarrow}$ for the spin direction $\sigma$ of interest
\begin{align}
	c_{s\sigma} = b_{\uparrow}c_{s\uparrow} + b_{\downarrow}c_{s\downarrow}.
\end{align}
For a spin pointing in the direction ($\theta,\varphi$), use
\begin{align}
	b_{\uparrow} =& \cos(\frac{\theta}{2}),\nonumber\\
	b_{\downarrow} =& \sin(\frac{\theta}{2})e^{-i\varphi}.
\end{align}
The coefficients for the opposite spin follows as
\begin{align}
	\bar{b}_{\uparrow}	=& b_{\downarrow}^{*},\nonumber\\
	\bar{b}_{\downarrow}	=& -b_{\uparrow}^{*}.
\end{align}

\subsubsection{Calculate relevant $a$'s in the chosen basis}
\begin{align}
	a_{s,\bar{\sigma},s,\sigma}^{\bar{\sigma}\sigma\sigma\sigma}	=& \sum_{\kappa\kappa'\kappa''\kappa'''}
		b_{\kappa}^{*}b_{\kappa'}\bar{b}_{\kappa''}b_{\kappa'''}^{*}\left(A_{s,\kappa'',s,\kappa'''}^{\kappa''\kappa\kappa\kappa'} + A_{s,\kappa''',s,\kappa''}^{\kappa'''\kappa'\kappa'\kappa*}\right),\nonumber\\
	a_{s+b,\sigma,s,\sigma}^{\sigma\sigma\sigma\sigma}
		=& \sum_{\kappa\kappa'\kappa''}
		b_{\kappa}^{*}|b_{\kappa'}|^2 b_{\kappa''}A_{s+b,\kappa'',s,\kappa'}^{\kappa''\kappa\kappa\kappa'},\nonumber\\
	a_{s+b,\bar{\sigma},s,\sigma}^{\bar{\sigma}\sigma\sigma\sigma}
		=& \sum_{\kappa\kappa'\kappa''}
		b_{\kappa}^{*}|b_{\kappa'}|^2 \bar{b}_{\kappa''}A_{s+b,\kappa'',s,\kappa'}^{\kappa''\kappa\kappa\kappa'}.
\end{align}

\subsubsection{Calculate sink-sources and currents}
\begin{align}
	\langle \hat{S}_{s}\rangle =& 2\textrm{Re}\left(a_{s,\bar{\sigma},s,\sigma}^{\bar{\sigma}\sigma\sigma\sigma}
	\sum_{\kappa\kappa'}	\bar{b}_{\kappa}^{*}b_{\kappa'}\langle c_{s\kappa}^{\dagger}c_{s\kappa'}\rangle_{H}\right),\nonumber\\
	\langle\hat{J}_{\sigma\sigma}^{b}\rangle =& -2\textrm{Re}\left(a_{s+b,\sigma,s,\sigma}^{\sigma\sigma\sigma\sigma} \sum_{\kappa\kappa'}b_{\kappa}^{*}b_{\kappa'}\langle c_{s+b,\kappa}^{\dagger}c_{s\kappa'}\rangle_{H}\right),\nonumber\\
	\langle\hat{J}_{\sigma\bar{\sigma}}^{b}\rangle =& -2\textrm{Re}\left(a_{s+b,\bar{\sigma},s,\sigma}^{\bar{\sigma}\sigma\sigma\sigma}\sum_{\kappa\kappa'}\bar{b}_{\kappa}^{*}b_{\kappa'}\langle c_{s+b,\kappa}^{\dagger}c_{s\kappa'}\rangle_{H}\right).
\end{align}

\subsubsection{Current vector field}
From the currents on the bonds, a current vector field, defined on site $s$, is finally calculated as
\begin{align}
	\mathbf{J}_{\sigma\sigma'}^{s} = \frac{1}{2}\sum_{b\neq 0}\mathbf{b}\langle\hat{J}_{\sigma\sigma'}^{b}\rangle
\end{align}

\subsection{Tabulating relevant $a$'s}
\label{Section:Tabulating_relevant_as}
Having summarized the method, it only remains to evaluate the $a$ coefficients in the basis of the Hamiltonian, which are denoted by a capital $A$.
By counting the number of spin indices in Eq.~\eqref{Equation:Definition_a} to be six, it may seem like we in general need to evaluate $2^6 = 64$ $a$'s per bond to be fully able to expand the commutators on the form in Eq.~\eqref{Equation:Bond_decomposition_reduced}.
However, the three $\delta$-functions ensure that only $2^3 = 8$ of these are non-zero.
For any given bond, we can therefore restrict ourselves to evaluate $a_{s+b,\lambda,s,\kappa'}^{\lambda\kappa\kappa\kappa'}$.
Further, Eq.~\eqref{Equation:Relations_aa} allows us to reduce this number to four, by fixing $\kappa' = \uparrow$.
The right hand side of Eq.~\eqref{Equation:Bond_decomposition_reduced} can then be completely determined for any site $s$ from a full knowledge of the table
\begin{center}
	\begin{tabular}{c|c|c|c|c}
		$(i,j)$	& $A_{i\uparrow j\uparrow}^{\uparrow\uparrow\uparrow\uparrow}$
				& $A_{i\uparrow j\uparrow}^{\uparrow\downarrow\downarrow\uparrow}$
				& $A_{i\downarrow j\uparrow}^{\downarrow\uparrow\uparrow\uparrow}$
				& $A_{i\downarrow j\uparrow}^{\downarrow\downarrow\downarrow\uparrow}$\\
				\hline
				\hline
				(1,1)&&&&\\
				\hline
				(1,2)&&&&\\
				\hline
				\vdots	& \vdots		& \vdots		& \vdots		& \vdots\\
				\hline
				($n$,$n$)&&&&
	\end{tabular}
\end{center}
Here $i,j$ are site indices, with $n$ the number of sites.
The elements necessary to reconstruct Eq.~\eqref{Equation:Bond_decomposition_reduced} for any given site $s$ can be found in the table by setting $i=s+b$ and $j=s$.
Note that also most terms in this table will be zero.
From Eq.~\eqref{Equation:Definition_a} it is clear that an $a$ will only be non-zero if the Hamiltonian contains some $f_{i\lambda j\lambda'}^{\alpha}$ connecting the two sites.
It is therefore sufficient to further limit the evaluation of the table to those pairs of sites that share a bond.

We also note that if we assume that each $h^{\alpha} = \sum_{\lambda\lambda'}h_{\lambda\lambda'}^{\alpha}$ is Hermitian, all results derived in this section applies to the individual $h^{\alpha}$'s.
In particular, the entries in the table for the full Hamiltonain $H$ can be obtained by adding the corresponding entries in the partial tables of the individual $h^{\alpha}$'s.

\subsection{Evaluating tables}
\label{Section:Evaluating_tables}
\subsubsection{Chemical potential}
The chemical potential can be written as
\begin{align}
	h^{\mu} =& \sum_{i\lambda}\mu_i c_{i\lambda}^{\dagger}c_{i\lambda} = \sum_{ij\lambda\lambda'}f_{i\lambda j\lambda}^{\mu}c_{i\lambda}^{\dagger}c_{i\lambda'}
\end{align}
where
\begin{align}
	f_{i\lambda j\lambda'}^{\mu} = \delta_{ij}\delta_{\lambda\lambda'}\mu_{i}.
\end{align}
The table is therefore (all entries are to be multiplied by $i/\hbar$)
\begin{center}
	\begin{tabular}{c|c|c|c|c}
		$(i,j)$	& $A_{i\uparrow j\uparrow}^{\uparrow\uparrow\uparrow\uparrow}$
				& $A_{i\uparrow j\uparrow}^{\uparrow\downarrow\downarrow\uparrow}$
				& $A_{i\downarrow j\uparrow}^{\downarrow\uparrow\uparrow\uparrow}$
				& $A_{i\downarrow j\uparrow}^{\downarrow\downarrow\downarrow\uparrow}$\\
				\hline
				\hline
				$(i,i)$	& $\mu_i$	& 0	& 0	& $\mu_i$\\
	\end{tabular}
\end{center}

\textit{Note: }
Although the chemical potential gives rise to some non-zero terms, it does not give rise to a contribution to any of the sink-source or current operators.
That it does not contribute to the currents is clear from the fact that it only connects terms on the same site.
To see that it also does not contribute to the sink-source term we note that the chemical potential term is rotationally invariant and can be considered to already be written in the $\sigma$-basis for any given $\sigma$.
That is, the $A$'s in the table can be considered to be written directly in the $\sigma$-basis.
Eq.~\eqref{Equation:Sink_source} reveals that only $a_{s,\eta,s,\eta'}^{\lambda\lambda'\kappa\kappa'}$ with $\lambda \neq \lambda'$ enters into the sink-source term, all of which are zero in the table above.

\subsubsection{Kinetic term}
The spin-degenerate kinetic energy can be written as
\begin{align}
	h^{t} =& \sum_{ij\lambda}t_{ij}c_{i\lambda}^{\dagger}c_{j\lambda} = \sum_{ij\lambda\lambda'}f_{i\lambda j\lambda'}^{t}c_{i\lambda}^{\dagger}c_{j\lambda'},
\end{align}
where 
\begin{align}
	f_{i\lambda j\lambda'}^{t} = \delta_{\lambda\lambda'}t_{ij}.
\end{align}
The table is therefore (all entries are to be multiplied by $i/\hbar$)
\begin{center}
	\begin{tabular}{c|c|c|c|c}
		$(i,j)$	& $A_{i\uparrow j\uparrow}^{\uparrow\uparrow\uparrow\uparrow}$
				& $A_{i\uparrow j\uparrow}^{\uparrow\downarrow\downarrow\uparrow}$
				& $A_{i\downarrow j\uparrow}^{\downarrow\uparrow\uparrow\uparrow}$
				& $A_{i\downarrow j\uparrow}^{\downarrow\downarrow\downarrow\uparrow}$\\
				\hline
				\hline
				$(i,j)$	& $t_{ij}$	& 0	& 0	& $t_{ij}$\\
	\end{tabular}
\end{center}

\subsubsection{Zeeman term}
The Zeeman term can be written as
\begin{align}
	h^{V_z} =& \sum_{i\lambda}V_z
		\left(\mathbf{\hat{n}}\cdot\boldsymbol{\sigma}\right)_{\lambda\lambda'}
		c_{i\lambda}^{\dagger}c_{i\lambda} = \sum_{ij\lambda\lambda'}f_{i\lambda j\lambda'}^{V_z}c_{i\lambda}^{\dagger}c_{j\lambda'},
\end{align}
where
\begin{align}
	f_{i\lambda j\lambda'}^{V_z} = \delta_{ij}\delta_{\lambda\lambda'}V_z
		\left(\mathbf{\hat{n}}\cdot\boldsymbol{\sigma}\right)_{\lambda\lambda'}.
\end{align}
The table is therefore (all entries are to be multiplied by $i/\hbar$)
\begin{center}
	\begin{tabular}{c|c|c|c|c}
		$(i,j)$	& $A_{i\uparrow j\uparrow}^{\uparrow\uparrow\uparrow\uparrow}$
				& $A_{i\uparrow j\uparrow}^{\uparrow\downarrow\downarrow\uparrow}$
				& $A_{i\downarrow j\uparrow}^{\downarrow\uparrow\uparrow\uparrow}$
				& $A_{i\downarrow j\uparrow}^{\downarrow\downarrow\downarrow\uparrow}$\\
				\hline
				\hline
				$(i,i)$	& $V_z n_z$	& $V_z\left(n_x - in_y\right)$	& $V_z\left(n_x + in_x\right)$	& $-V_z n_z$\\
	\end{tabular}
\end{center}

\subsubsection{Rashba spin-orbit interaction}
The Rashba spin-orbit interaction can be written as
\begin{align}
	h^{SO} =& \sum_{ij\lambda\lambda'}\alpha_{ij}\left(\left(\delta_{i,j-\hat{x}} - \delta_{i,j+\hat{x}}\right)\left(-i\sigma_{y}\right)_{\lambda\lambda'}\right.\nonumber\\
	&\left. + i\left(\delta_{i,j-\hat{y}} - \delta_{i,j+\hat{y}}\right)\left(\sigma_{x}\right)_{\lambda\lambda'}\right)
	c_{i\lambda}^{\dagger}c_{j\lambda'}\nonumber\\
	=& \sum_{ij\lambda\lambda'}f_{i\lambda j\lambda'}^{SO}c_{i\lambda}^{\dagger}c_{j\lambda'},
\end{align}
where
\begin{align}
	f_{i\lambda j\lambda'}^{SO} =& \alpha_{ij}\left(\left(\delta_{i,j-\hat{x}} - \delta_{i,j+\hat{x}}\right)\left(-i\sigma_{y}\right)_{\lambda\lambda'}\right.\nonumber\\
	&\left. + i\left(\delta_{i,j-\hat{y}} - \delta_{i,j+\hat{y}}\right)\left(\sigma_{x}\right)_{\lambda\lambda'}\right).
\end{align}
The table is therefore (all entries are to be multiplied by $i/\hbar$)
\begin{center}
	\begin{tabular}{c|c|c|c|c}
		$(i,j)$	& $A_{i\uparrow j\uparrow}^{\uparrow\uparrow\uparrow\uparrow}$
				& $A_{i\uparrow j\uparrow}^{\uparrow\downarrow\downarrow\uparrow}$
				& $A_{i\downarrow j\uparrow}^{\downarrow\uparrow\uparrow\uparrow}$
				& $A_{i\downarrow j\uparrow}^{\downarrow\downarrow\downarrow\uparrow}$\\
				\hline
				\hline
				$(i,i+\hat{x})$	& 0	& $-\alpha_{i,i+\hat{x}}$	& $\alpha_{i,i+\hat{x}}$	& 0\\
				\hline
				$(i,i-\hat{x})$	& 0	& $\alpha_{i,i-\hat{x}}$	& $-\alpha_{i,i-\hat{x}}$	& 0\\
				\hline
				$(i,i+\hat{y})$	& 0	& $i\alpha_{i,i+\hat{y}}$	& $i\alpha_{i,i+\hat{y}}$	& 0\\
				\hline
				$(i,i-\hat{y})$	& 0	& $-i\alpha_{i,i-\hat{y}}$	& $-i\alpha_{i,i-\hat{y}}$	& 0\\
	\end{tabular}
\end{center}

\subsubsection{Total}
For a total Hamiltonain of the form
\begin{align}
	H = H^{\mu} + H^{t} + H^{V_z} + H^{SO},
\end{align}
the total table is obtanied as the sum of the individual tables (all entries are to be multiplied by $i/\hbar$)
\begin{center}
	\begin{tabular}{c|c|c|c|c}
		$(i,j)$	& $A_{i\uparrow j\uparrow}^{\uparrow\uparrow\uparrow\uparrow}$
				& $A_{i\uparrow j\uparrow}^{\uparrow\downarrow\downarrow\uparrow}$
				& $A_{i\downarrow j\uparrow}^{\downarrow\uparrow\uparrow\uparrow}$
				& $A_{i\downarrow j\uparrow}^{\downarrow\downarrow\downarrow\uparrow}$\\
				\hline
				\hline
				$(s,s)$			& $V_z n_z$	& $V_z\left(n_x - in_y\right)$	& $V_z\left(n_x + in_y\right)$	& $-V_z n_z$\\
				\hline
				$(s-\hat{x},s)$	& $t_{s-\hat{x},s}$	& $-\alpha_{s-\hat{x},s}$	& $\alpha_{s-\hat{x},s}$	& $t_{s-\hat{x},s}$\\
				\hline
				$(s+\hat{x},s)$	& $t_{s+\hat{x},s}$	& $\alpha_{s+\hat{x},s}$	& $-\alpha_{s+\hat{x},s}$	& $t_{s+\hat{x},s}$\\
				\hline
				$(s-\hat{y},s)$	& $t_{s-\hat{y},s}$	& $i\alpha_{s-\hat{y},s}$	& $i\alpha_{s-\hat{y},s}$	& $t_{s-\hat{y},s}$\\
				\hline
				$(s+\hat{y},s)$	& $t_{s+\hat{y},s}$	& $-i\alpha_{s+\hat{y},s}$	& $-i\alpha_{s+\hat{y},s}$	& $t_{s+\hat{y},s}$\\
	\end{tabular}
\end{center}
We have here ignored the contributions from $H^{\mu}$, because as noted above all relevant terms are zero.
We have also assumed that $t_{ij}$ only is non-zero for nearest neighbour terms.

\subsection{Note on superconductivity}
\label{Section:Note_on_superconductivity}
All terms in the Hamiltonian treated so far are quadratic of the form $c_{i\lambda}^{\dagger}c_{j\lambda}$.
However, even within a mean-field treatment of ordinary $s$-wave superconductivity, we will also encounter terms in the Hamiltonian which are of the form
\begin{align}
	h^{SC} = \sum_{i}\Delta_i c_{i\uparrow}^{\dagger}c_{i\downarrow}^{\dagger} + \Delta^{*}c_{i\downarrow}c_{i\uparrow},
\end{align}
where $\Delta_i = -V_{sc}\langle c_{i\downarrow}c_{i\uparrow}\rangle$, for some $V_{sc}$.
This term is a spin-singlet and therefore appears similarly in any given spin-basis, such that we can assume that $\sigma = \uparrow$.
If the term is included in the Hamiltonian, then it is clear from
\begin{align}
	\frac{i}{\hbar}\left[h^{SC}, c_{s\uparrow}^{\dagger}c_{s\uparrow}\right] =& \frac{i}{\hbar}\left(\Delta^{*}c_{s\downarrow}c_{s\uparrow} - \Delta_{s}c_{s\uparrow}^{\dagger}c_{s\downarrow}^{\dagger}\right)
\end{align}
that Eq.~\eqref{Equation:Density_source_current} has to be modified to read
\begin{align}
	\label{Equation:Superconductivity_commutator}
	\frac{d\hat{\rho}_{s\uparrow}}{dt} =& \hat{S}_{s} - \sum_{b}\left(\hat{J}_{\uparrow\uparrow}^{b} + \hat{J}_{\uparrow\downarrow}^{b}\right) + \frac{i}{\hbar}\left(\Delta_{s}^{*}c_{s\downarrow}c_{s\uparrow} - \Delta_{s}c_{s\uparrow}^{\dagger}c_{s\downarrow}^{\dagger}\right).
\end{align}
Taking the expectation value of this expression and using that $\Delta_s = -V_{sc}\langle c_{s\downarrow}c_{s\uparrow}\rangle = \left(-V_{sc}\langle c_{s\uparrow}^{\dagger}c_{s\downarrow}^{\dagger}\rangle\right)^{*} = \left(\Delta_{s}^{*}\right)^{*}$ we find
\begin{align}
	\frac{d\langle\hat{\rho}_{s\uparrow}\rangle}{dt} =& \langle\hat{S}_{s}\rangle - \sum_{b}\left(\langle\hat{J}_{\uparrow\uparrow}^{b}\rangle + \langle\hat{J}_{\uparrow\downarrow}^{b}\rangle\right).
\end{align}
Rewriting this as
\begin{align}
	d\langle\hat{\rho}_{s\uparrow}\rangle =& \left(\langle\hat{S}_{s}\rangle - \sum_{b}\left(\langle\hat{J}_{\uparrow\uparrow}^{b}\rangle + \langle\hat{J}_{\uparrow\downarrow}^{b}\rangle\right)\right)dt,
\end{align}
it is possible to view any non-zero term as a possible generator of density fluctuations.
Although the superconducting condensate can carry current, it is clear from the vanishing expectation value that it does not in itself generate density fluctuations.
In fact, it is clear from Eq.~\eqref{Equation:Superconductivity_commutator} that the superconducting term is unable to generate any currents at all, including currents which preserves the density, as it gives rise to a purely on-site contribution to $d\hat{\rho}_{s\uparrow}/dt$.
The superconducting term is therefore not responsible for generating any currents; the condensate only carries them once they exist.
The treatment of only ordinary quadratic terms is therefore sufficient for calculating the current also in the presence of $s$-wave superconductivity.

\end{document}